\documentclass[twocolumn]{revtex4-1}

\usepackage{graphicx,dcolumn,bm,amsmath}
\usepackage{color}

\begin{document}
\title[Representational Analysis of Atomistic Ensembles]{Representational Analysis of Extended Disorder in Atomistic Ensembles Derived from Total Scattering Data}

\author{James~R.~Neilson}
\address{Department of Chemistry, Colorado State University, CO}
\email{james.neilson@colostate.edu}
\author{Tyrel~M.~McQueen}
\address{Department of Chemistry, Department of Materials Science and Engineering, and Department of Physics \& Astronomy, Johns Hopkins University, Baltimore, MD}
\email{mcqueen@jhu.edu}

\begin{abstract}
With the increased availability of high intensity time-of-flight neutron and synchrotron X-ray scattering sources that can access wide ranges of momentum transfer, the pair distribution function method has become a standard analysis technique for studying disorder of local coordination spheres and at intermediate atomic separations.  In some cases, rational modeling of the total scattering data (Bragg and diffuse) becomes intractable with least-squares approaches and necessitates reverse Monte Carlo (RMC) simulations using large atomistic ensembles.  However, the extraction of meaningful information from the resulting atomistic ensembles is challenging, especially at intermediate length scales.  We use representational analysis to describe displacements of atoms in  RMC ensembles from an ideal crystallographic structure in an approach analogous to tight-binding methods.  Rewriting the displacements in terms of a local basis that is descriptive of the ideal crystallographic symmetry provides a robust approach to characterizing medium-range order (and disorder) and symmetry breaking in complex and disordered crystalline materials.  This method enables the extraction of statistically relevant displacement modes (orientation, amplitude, and distribution) of the crystalline disorder and provides directly meaningful information in a locally symmetry-adapted basis set that is most descriptive of the crystal chemistry and physics.  
\end{abstract}

\maketitle

\section{Introduction}
Achieving an atomistic description of solids continues to provide a challenge to the study of materials, especially as we learn that imperfections and disorder of crystals can give rise to the emergence of unexpected materials properties.  For example, the multifunctional properties of the perovskite manganites can only be explained by understanding the relationships between the local and average structures \cite{Bozin:2007uz,PhysRevB.76.174210}.  Therefore, we strive to further classify and quantify the nature of any local ordering (short-range order) that is patterned in a disordered fashion.  Pair distribution function analysis of total scattering data has become a common technique for the characterization of local distortions and disorder in crystals, as well as nanoparticle structures \cite{Egami_Billinge,Billinge27042007,young2011applications,keen2015crystallography}.  

Modeling of atomistic structures -- with emphasis on capturing the correct \emph{local structure}  -- from experimentally derived atom-atom histograms poses a great challenge, especially when the best description of the PDF has a short, finite correlation length (a domain) that becomes averaged into a higher symmetry in the crystallographic structure.  To obtain an atomistic description of such a model with these domains (where each domain consists of a few unit cells), simulations containing thousands of atoms can be used to model the total scattering data.  By employing a large-scale simulation, the limitations from periodic boundary conditions are lifted, thus allowing disordered aspects of the structure to either average out into the Debye-Waller factor in the case of crystalline disorder or lack any attributes of long-range order over the range of data provided in reciprocal space (after convolution with the finite size of the simulation) in order to describe amorphous solids \cite{Renninger19741,McGreevy:1988bu,Elliot_Physics}.  However, analysis of these large-scale atomistic ensembles containing 1000's of atoms has been non-trivial, both in the challenge of extracting information relative to the average crystallographic structure and also in providing statistically meaningful information; there are typically many more free parameters in these simulations than independent observations (i.e., data).    

\begin{figure}[t]
\begin{center}
\includegraphics[width=3.in]{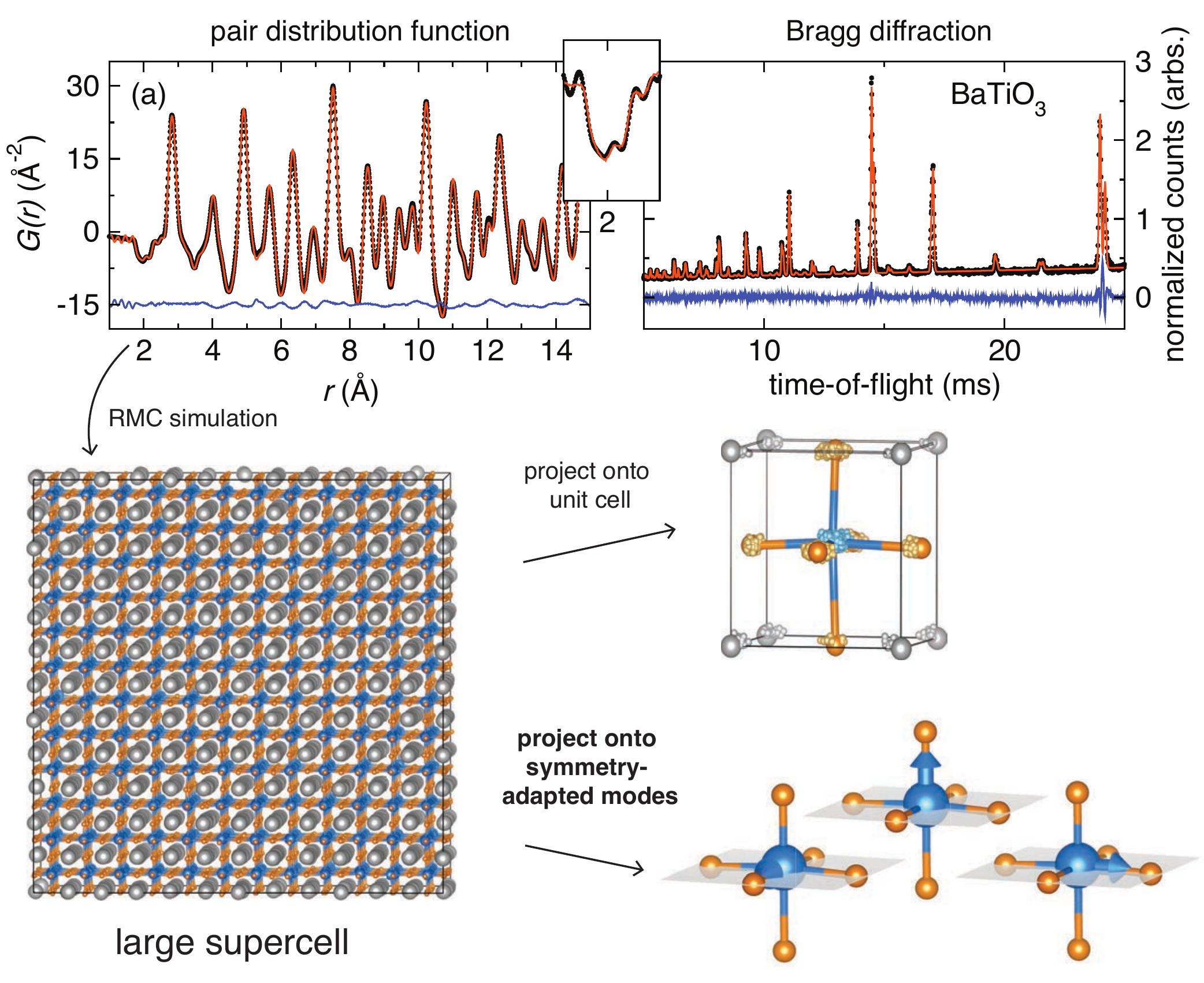}
\caption{Schematic illustration of the workflow for generating a large ensemble simulation using reverse Monte Carlo simulations, which can be folded back onto a small unit cell or the atomic displacements can be projected onto the tight-binding modes of small unit cell.  }
\label{fig:outline}
\end{center}
\end{figure}

Herein, we develop a systematic approach for analyzing the disorder in large atomistic simulations of complex crystal structures using representational analysis.  The determination of crystallographic superstructures resulting from displacive distortions via symmetry-mode analysis of a statistical distribution of ensembles has proven to be very powerful (cf., WO$_3$ and LaMnO$_3$) \cite{kerman2012superstructure}.  Another similar approach, but coupled to a different analysis, has also made it possible to extract phonon dispersions from powder diffraction data \cite{PhysRevB.60.6204,PhysRevLett.93.075502,PhysRevB.72.214304}.  Here, we use a variation of this technique adapted to understand local structural variations by projecting displacements of atoms from their average crystallographic sites in atomistic ensembles onto a tight-binding-like basis formed from the symmetry-adapted\footnote{We define the modes as being \emph{locally} symmetry-adapted, since the symmetry relationship that we use is only strictly defined for $k=(0,0,0)$.  In a single unit cell, these modes are fully symmetry adapted} modes of a single unit cell, as depicted in Figure~\ref{fig:outline} -- we define these modes as \emph{tight-binding modes}.  When displacements from an ideal crystallographic site are projected onto this \emph{locally} symmetry-adapted basis, the disorder can be quantified and statistically analyzed to determine the frequency of specific displacement magnitudes and orientations.  This manuscript outlines the analytical method and presents two illustrative applications of the method: the observation of a trigonal distortion in BaTiO$_3$ at room temperature and the identification of the local displacement modes in the charge-ice pyrochlore, Bi$_2$Ti$_2$ O$_7$.  More broadly, our approach is equally important for the analysis of experimental diffraction and scattering data \cite{Shoemaker_PRB_2010,Shoemaker:2010ia,king2011high}, \emph{ab initio} and force-field-based simulations \cite{dixon2014origin,palin2014computer}, and combinations of the two \cite{B922993K,doi:10.1021/jp911108d}.  Furthermore, this approach provides a common language and representation for bridging experiment- and theory-derived models.

\section{Method}

\subsection{Introduction to Total Scattering Methods}

The analytical method described here operates on an ensemble of atoms that can be described as a enlarged big-box generated from small crystallographic unit cells.  The atom positions need not sit on precisely ordered lattice sites; however, upon back-folding the big-box ensemble onto the parent unit cell, the average atom positions should project close to particular lattice sites, each with a position distribution resembling something like a Debye-Waller factor (\emph{i.e.}, the model may be paracrystalline).  This method is agnostic to how the models are generated; the authors refer the reader to Refs.~\cite{Tucker:2007vt,Tucker:th0053,young2011applications,keen2015crystallography,Egami_Billinge} for a description of modeling total scattering data.  

Here, we use ``total scattering'' to refer to the scattering of X-rays or neutrons that describes the structure factor of the crystallographic symmetry (diffraction from periodically ordered components) and the diffuse scattering that can arise from displacements of atoms from their ideal lattice points, including displacements from thermal motion and static disorder in the crystal \cite{Egami_Billinge}.   If the total scattering structure factor, $S(Q)$, is measured to sufficiently high momentum transfer ($Q_{max} \gtrsim 15$ \AA$^{-1}$), one can numerically take a sine Fourier transform to convert $S(Q)$ into the reduced pair distribution function (PDF), $G(r)$, 
\begin{align}
G(r) = & 4 \pi r \rho_0 [g(r)-1] \\ \nonumber
= & \frac{2}{\pi} \int_{0}^{\infty} Q[S(Q)-1]\sin(Qr)dQ,
\end{align}
where $\rho_0$ is the average number density of the material and $g(r)$ is the atomic pair distribution function.  The atomic pair distribution function, $g(r)$, is a direct measure of the relative positions of atoms in a solid, that is, an experimentally-accessible real-space histogram of all atom-atom separations in the solid (of periodically ordered \emph{and} disordered atoms).  Because of the crystallographic phase problem, without the use of isotopic labeling or anomalous scattering, it is not possible to directly assign peaks in the PDF to specific atoms; therefore, atomic-scale modeling must be used to make assignments to individual peaks.

``Small-box'' models, which allow extraction of bond lengths and a description of the thermal motion (i.e., Debye-Waller factors), can be obtained from least-squares (LS) optimization of a crystallographic unit cell, or some small variant thereof to the experimental PDF  using the software, {\sc PDFgui}.\cite{Egami_Billinge,Proffen_1999,PDFgui}  LS optimization is susceptible to finding local minima in the goodness-of-fit and is numerically cumbersome when the model contains many degrees of freedom, as applicable here.  Additionally, these short-range ordered models often fail to provide an accurate description of the crystallographic observations \cite{Neilson_PRB_2012,PhysRevB.87.045124,king2013local}.

A complimentary approach to extract atomistic configurations from the PDF is to simultaneously model both the crystallographic structure factor and the PDF by employing a ``large-box''simulation of the total scattering data.  A Reverse Monte Carlo (RMC) algorithm can be used to find atomistic configurations of the ensemble consistent with both the experimentally determined $G(r)$ and $S(Q)$ \cite{Tucker:2007vt,Tucker:th0053}.

\subsection{Coordinate Transform and Decomposition}

\begin{figure}[t]
\begin{center}
\includegraphics[width=3.in]{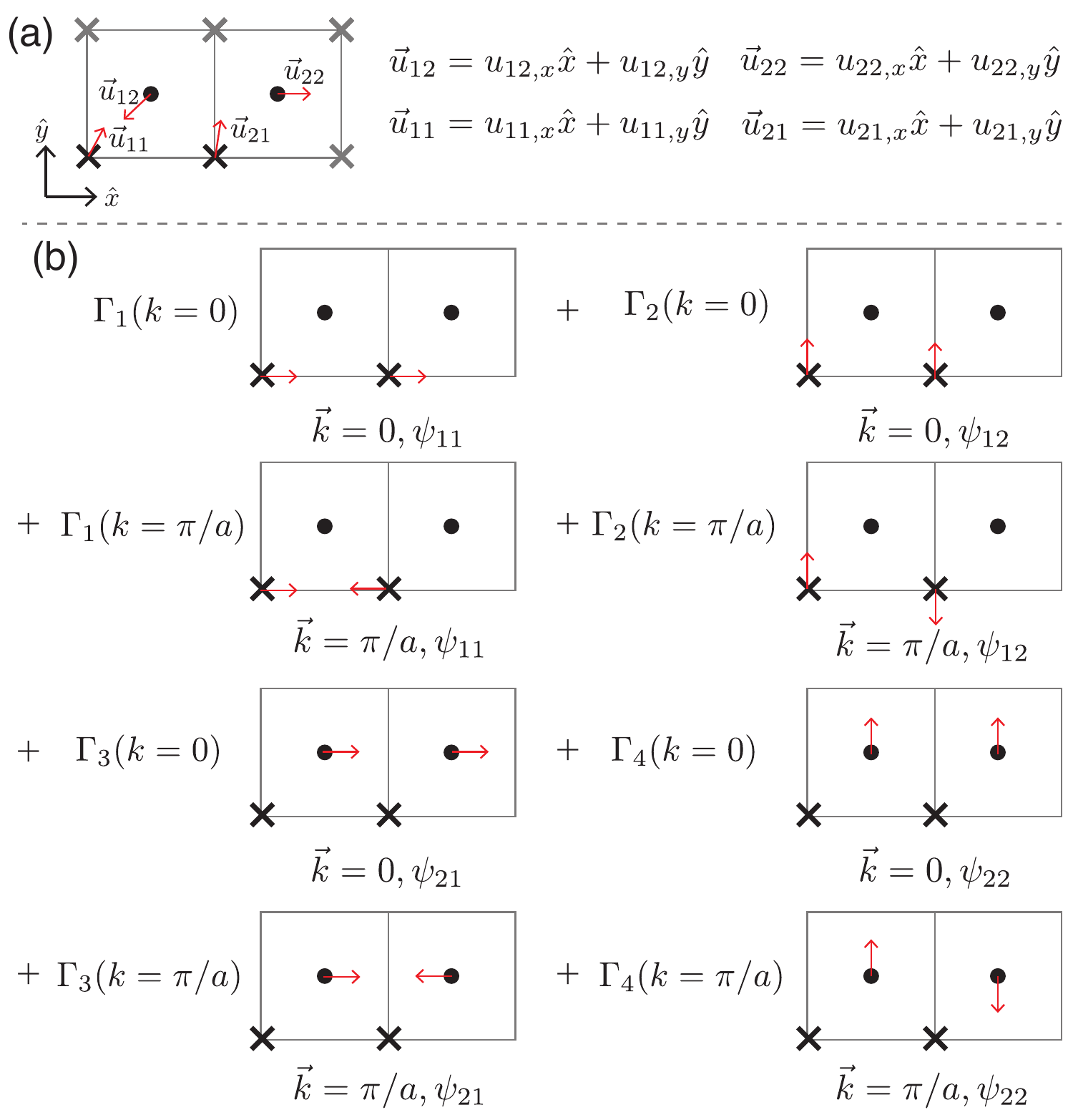}
\caption{Schematic illustration of the coordinate transformation from atomic displacements into tight-binding basis based on locally symmetry-adapted modes.  }
\label{fig:transform}
\end{center}
\end{figure}

The goal of this method is to define the atomic configurations of an ensemble as displacements from the ideal crystallographic positions.  Here, we define a ``large-box'' as a  $M_x \times M_y \times M_z$ enlargement of the crystallographic unit cell to form an atomistic ensemble, but no attempt is made to constrain the symmetry between atoms, either within the subcells or the ``large-box''.  The simplest such basis is to simply write down displacement vectors, in Cartesian or lattice coordinates, for each atom in the ensemble.  Each atom within the crystallographic unit cell, $i$, has a unique position defined by a vector, $\vec{x}_{i,n}$.  The vector, $\vec{R_n}$, describes the spatial vector between each unit cell, $n$, within the ensemble.   Each atom can be mapped as a displacement from its ideal  position in the crystallographic unit cell, $\vec{x}_{i,n}^0$, by $\vec{u}_{i,n} = \vec{x}_{i,n} - \vec{x}_{i,n}^0$, where the values $\vec{x}_{i,n}^0$ are often determined from a traditional crystallographic analysis (Rietveld analysis or single-crystal structural refinement).  Such a representation is shown schematically for a simple two dimensional toy model, a $2\times1$ ``big-box'' built from a crystallographic unit cell with two atoms and $C_4$ symmetry, in Figure~\ref{fig:transform}(a).  While straightforward to compute, this basis (the displacement vectors) lacks any connection to the symmetries that are present, locally or on average, and are thus difficult to interpret.  A more refined approach is to rewrite the displacements in terms of the normal modes of the crystallographic structure, with amplitudes and phases for every mode at every wavevector in the Brillouin zone (as determined by the point symmetry of each wavevector).  This normal-mode basis provides physical insight because the atomic displacements are mapped onto symmetry-defined motions away from their ideal positions, and correlations between unit cells are captured.  There is, however, an even better choice of basis that keeps many of the advantages of the classic normal mode approach but retains physical insight into the \emph{local} symmetry changes.  

First, the local tight-binding (\emph{i.e.}, locally symmetry-adapted) modes are identified.  This is accomplished by rewriting all possible atomic motions within a single unit cell into motions consistent with the point symmetry of the crystal at the Brillouin zone center, $k = (0,0,0)$.  Each motion (or mode) can be labelled according to the irreducible representation that it transforms under in the point symmetry group and is described by a set of basis vectors describing the actual atomic motions.  Identification of these tight-binding modes is straightforward: basis vectors spanning each irreducible representations for each space group have been tabulated by Kovalev \cite{kovalev}, or can be computed by various crystallographic tools, including {\sc karep} \cite{KAREP}, {\sc sara}$h$ \cite{sarah}, {\sc basireps} \cite{basireps}, the Bilbao Crystallographic Server (Symmetry Adapted Modes)\cite{aroyo2011crystallography,aroyo2006bilbao,aroyo2006bilbao2,doi:10.1080/0141159031000076110}, or the ISOTROPY software suite \cite{isotropy}.  The input for these tools are the crystallographic space group and atom positions of the small (crystallographic average) unit cell as one would derive from Rietveld (or other suitable crystallographic) analysis.

These tight-binding modes provide an orthonormal and \emph{local} basis for describing all possible motional degrees of freedom within a single unit cell, and are analogous to normal vibrational modes of a molecular system.  To retain this physical intuition but capture the degrees of freedom of a ``large-box'' atomistic ensemble, we adopt a technique analogous to tight-binding methods in electronic structure calculations \cite{PhysRev.94.1498} and write down modes at a  non-zero wavevector in terms of these local basis functions that we define at the Brillouin zone center (a la atomic orbitals), with appropriate phase factors to describe correlations between crystallographic unit cells in an ensemble.  Specifically, we define the spatial correlations between unit cells within the ensemble with a quantized, reciprocal wavevector, $\vec{k} = 2\pi/\vec{R}$.   The vector spans the indices, $\vec{k_x} = (2\pi n_x/M_x, 2\pi n_y/M_y, 2\pi n_z/M_z)$ for all $n_x = 0,1, \cdots (M_x-1)$, $n_y = 0, 1, \cdots (M_y-1)$, and $n_z = 0, 1, \cdots (M_z-1)$; in other words, the wavevectors are in steps of $2\pi/M$ along each direction.  For mathematical convenience, we define all values for $\vec{k}$ as positive.  
The amplitude of a tight-binding mode,  $\Gamma_{j,\lambda}(\vec{k})$, with the associated phase factor described by the reciprocal-space wavevector, $\vec{k}$, is defined by: 
\begin{equation}
\Gamma_{j,\lambda}(\vec{k}) = \sum_{n} \sum_{i} \vec{u}_{i,n} \cdot \vec{\psi}_{i,j,\lambda} \exp(-i  \vec{k} \cdot \vec{R_n}~),  \label{eqn:irrep}
\end{equation}
 where $i$ runs over \emph{all} atoms in the crystallographic unit cell, and $n$ runs over all unit cells contained within the ensemble.  The vector, $\vec{R}$, points to the $n^{th}$ crystallographic unit cell in the ensemble.  The values, $\vec{\psi}_{i,j,\lambda}$, are the vectorial contribution of atom $i$ to the mode described by the ($j,\lambda$) pair.   The vectorial contributions can span multiple atoms, as pertaining to the crystallographic multiplicity of the particular site in the original crystallographic unit cell.  The index, $j$, specifies each set of modes that together transform as an irreducible representation of the point group; $\lambda$ is equal to the dimensionality of the corresponding irreducible representation and runs over all modes in the set.  Together, there are $3N$ distinct $j,\lambda$ pairs, or tightly-binding  modes, where $N$ is the number of atoms in the small, crystallographic cell.

 There is no index $k$ on $\vec{\psi}$, just like there is no wavevector dependence to atomic orbitals in the classic tight-binding electronic structure approach, because all wavevector dependences are explicitly included in the phase factors.  Further, note that to retain all degrees of freedom we allow the amplitudes of each tight-binding mode  to be independent of all others, even if symmetry would constrain them (i.e., because one irreducible representation may be spanned by multiple modes). This allows for us to consider, but not enforce, symmetry in describing the ``large-box'' atomistic ensembles.  Stated differently, the projection is only a change of basis; all $3N-6$ degrees of freedom (for an ensemble of $N$ atoms) are retained (omitting the 3 translational and 3 rotational degrees of freedom), and the exact atomistic ensemble can be reconstructed by the inverse of \label{eqn:irrep}:

\begin{equation}
\vec{u}_{i,n} = \sum_{\vec{k}} \sum_{j} \sum_{\lambda}\Gamma_{j,\lambda}(\vec{k}) \cdot \vec{\psi}_{i,j,\lambda} \exp(i  \vec{k} \cdot \vec{R_n}~).  \label{eqn:rebuild}
\end{equation}

This method, as applied to the toy model, is shown in Figure~\ref{fig:transform}(b).  We note that this is distinct from typical crystallographic order parameter analysis  \cite{PhysRevLett.93.075502,PhysRevB.72.214304,PhysRevB.60.6204,isodisplace,isotropy,kerman2012superstructure}, 
in which the constraints of the parent crystallographic symmetry are preserved  and the primary interoperable variables are the order parameter amplitudes, thus providing one number for a pair of basis vectors that describe a displacement transforming as a two-dimensional irreducible representation, vs.\ two numbers in our approach.  We retain all possible degrees of freedom.

\subsection{Continuous symmetry measures}

When using our tight-binding modes, we can determine the  activity of the mode and deviation of the ensemble from the crystallographic symmetry not just from the mode amplitude, but also from its mean-squared deviation (MSD) from an ensemble operated on by a symmetry operation of the parent crystallographic space group.  There are at least two distinct types of continuous symmetry measures (as developed by D. Avnir and coworkers \cite{zabrodsky1992continuous,alvarez2005continuous}) that we characterize here.  
First, the global activity of a single tight-binding mode ($j$, consisting of one or more individual $\lambda$ modes depending on the dimensionality of the corresponding irreducible representation) can be quantified as the mean-squared deviation (MSD) between the ($|\Gamma_{j,\lambda}(\vec{k})|$) amplitudes and the new amplitude coefficients, $|\Gamma_{G,j,\lambda}(\vec{k})^\prime|$, following application of a symmetry operation, $G$, of the crystallographic space group:
\begin{equation}
s_{G,j} = \Big( \displaystyle\sum\limits_{\vec{k}} (\displaystyle\sum\limits_{\lambda}|\Gamma_{G,j,\lambda}(\vec{k})^\prime|^2 - \displaystyle\sum\limits_{\lambda}|\Gamma_{j,\lambda}(\vec{k})|^2)^2 \Big)^{1/2}.  \label{eqn:csmirrep}
\end{equation}
For purely symmetry-conserving displacements, the MSD should be zero. Here, it is critical to combine the squared amplitudes of all individual modes that together transform as a single multidimensional irreducible representation (the innermost sums) because the amplitude of individual modes can be varied simply by changing the choice of basis vectors within that mode set. The sum over all wavevectors is justified to identify local symmetry changes because it corresponds to summing up contributions derived from a single local tight-binding mode (a la atomic orbital), and is exact in the molecular limit. The final square root is provided for convenience to make the magnitude of $s_{G,j}$ more physically interpretable.

The related MSD, not broken down by individual mode sets, is similarly simple to calculate:
\begin{equation}
s_{G} = \Big( \displaystyle\sum\limits_{j}\displaystyle\sum\limits_{\vec{k}} (\displaystyle\sum\limits_{\lambda}|\Gamma_{G,j,\lambda}(\vec{k})^\prime|^2 - \displaystyle\sum\limits_{\lambda}|\Gamma_{j,\lambda}(\vec{k})|^2)^2 \Big)^{1/2}.  \label{eqn:csm}
\end{equation}
Where again the final square root is provided for convenience.

\begin{figure}[t]
\begin{center}
\includegraphics[width=3.in]{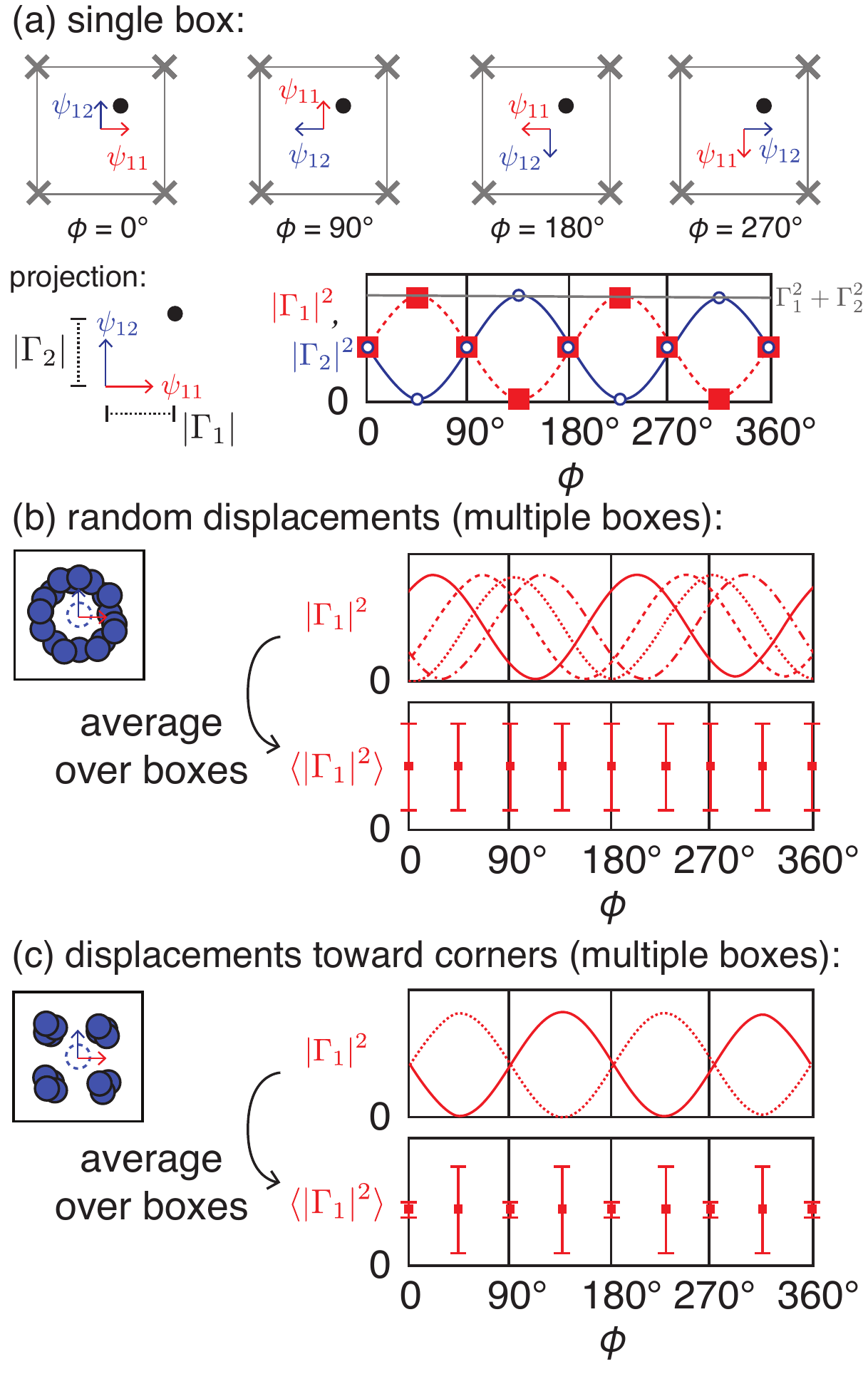}
\caption{Schematic illustration of the the rotation of the Euler angle about the rotation axis of a particular symmetry element ($C_4$).  (a) In a single box, the amplitudes of the displacements oscillate as a function of Euler angle, $\phi$.  (b) For random displacements across multiple subcells or boxes, the amplitude averages to a constant value as a function of $\phi$ with a high and constant variance (denoted by error bars).  (c) For displacements towards the four corners, the amplitude averages to a constant value as a function of $\phi$; however, the variance oscillates as a function of $\phi$ angle.   }
\label{fig:SchemeEuler}
\end{center}
\end{figure}

The second type of deviations from the parent space group that can be identified are distortions that do not retain an equivalence of mode amplitudes within a single mode set that transforms as a multidimensional irreducible representation.  To illustrate this, consider a single box with $C_4$ symmetry and an atom in the center displaced along the diagonal direction (Figure~\ref{fig:SchemeEuler}). Projected onto the two basis vectors $\Gamma_1$ and $\Gamma_2$, which together span the two-dimensional,  $E$, irreducible representation in the corresponding point group, the amplitudes along each basis are initially equal. As the Euler angle that defines the absolute orientation of the basis vectors is varied, the intensity of $\Gamma_2$ reduces while $\Gamma_1$ rises until $\Gamma_1$ is collinear with the atom displacement; this oscillatory pattern continues the rest of the way. Note that the sum of the square amplitudes from the two contributions ($\Gamma_1$ and $\Gamma_2$) is a constant (this is required as the magnitude of the displacement is not changing).   However, across multiple subcells or across multiple simulation runs, one can differentiate between random and ordered displacements.  Let $\phi_0$ be the initial angle of the displacement of the center atom. Different $\phi_0$'s correspond to phase shifts of the values of $\Gamma_1^2$ (and $\Gamma_2^2$).  If the $\phi_0$'s are completely random, then their average is a flat line as a function of Euler angle, with variances that are also flat [Figure~\ref{fig:SchemeEuler}(b)]. On the other hand, if the $\phi_0$'s are pinned to specific directions, then only a subset of the phase shifts are present. This will often result in an average that is still flat as a function of Euler angle, \emph{e.g.}, if they are pinned every 90$^\circ$, but the variances will no longer be uniform [Figure~\ref{fig:SchemeEuler}(c)].   This can be leveraged to determine whether the displacements are approximately random or fixed in some subset of orientations relative to the parent unit cell coordinate system.

\section{Case Studies}

\subsection{Trigonal displacements in tetragonal BaTiO$_3$}

\begin{figure}[t]
\begin{center}
\includegraphics[width=2.75in]{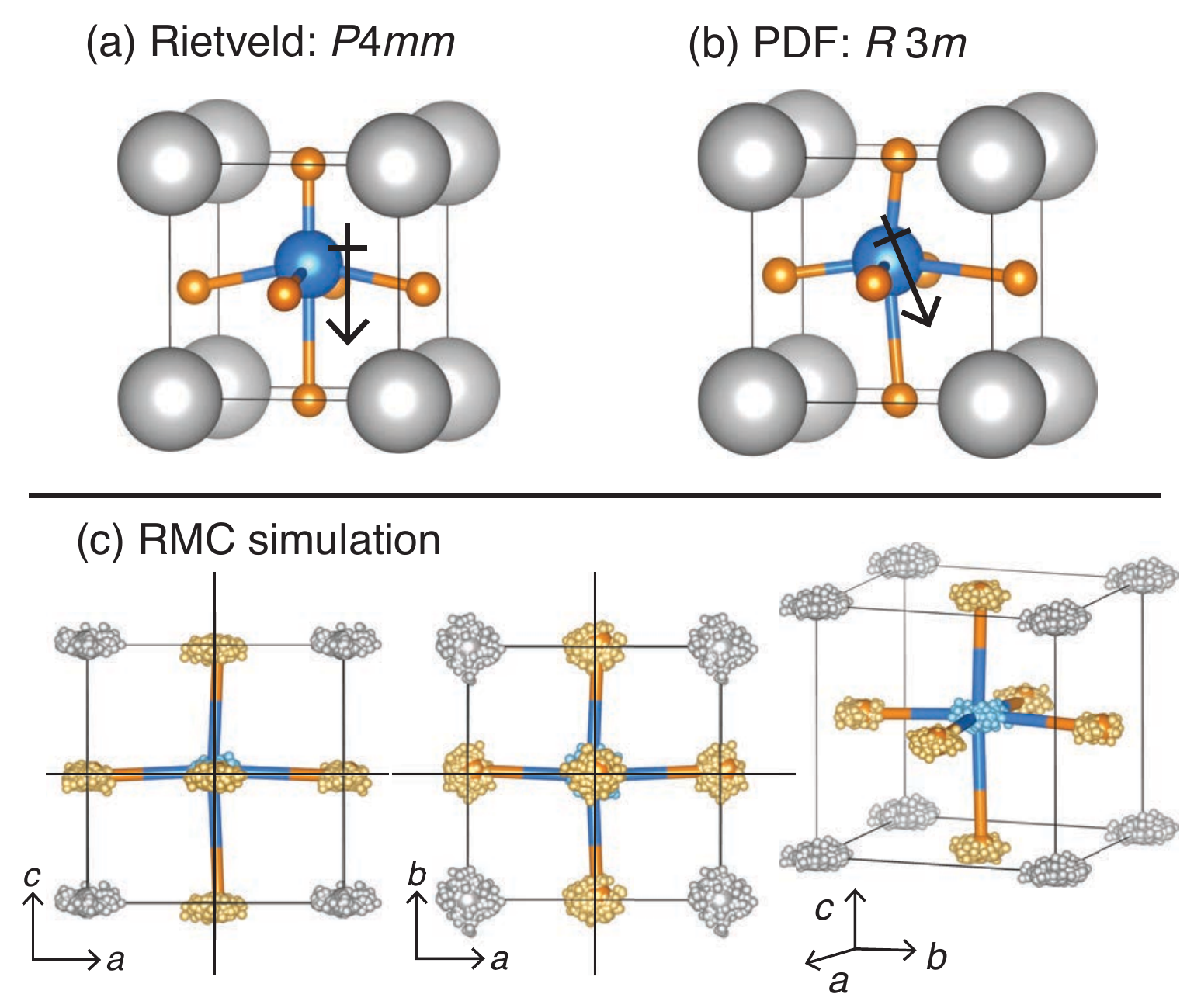}
\caption{
(a) Room temperature $P4mm$ crystal symmetry of BaTiO$_3$ with  exaggerated displacements of the Ti and O positions to illustrate the ferroelectric dipole.  (b) the pair distribution function analysis reveals a local distortion present at room temperature that resembles the low-temperature $R3m$ crystal structure \cite{kwei1995pair,ravel1998local,doi:10.1021/cm100440p}, illustrated here with  exaggerated Ti and O displacements.  (c) Folded atomistic big-box ensemble generated from a RMC simulation, overlain with the anisotropic thermal ellipsoids determined from small-box modeling of the PDF ($R3m$ structure).  The (200), (020), and (002) planes are shown to illustrate the net displacement of the oxygen atom positions rather than the titanium and barium atom positions.  }
\label{fig:BaTiO3clouds}
\end{center}
\end{figure}

\begin{figure}[t]
\begin{center}
\includegraphics[width=2.75in]{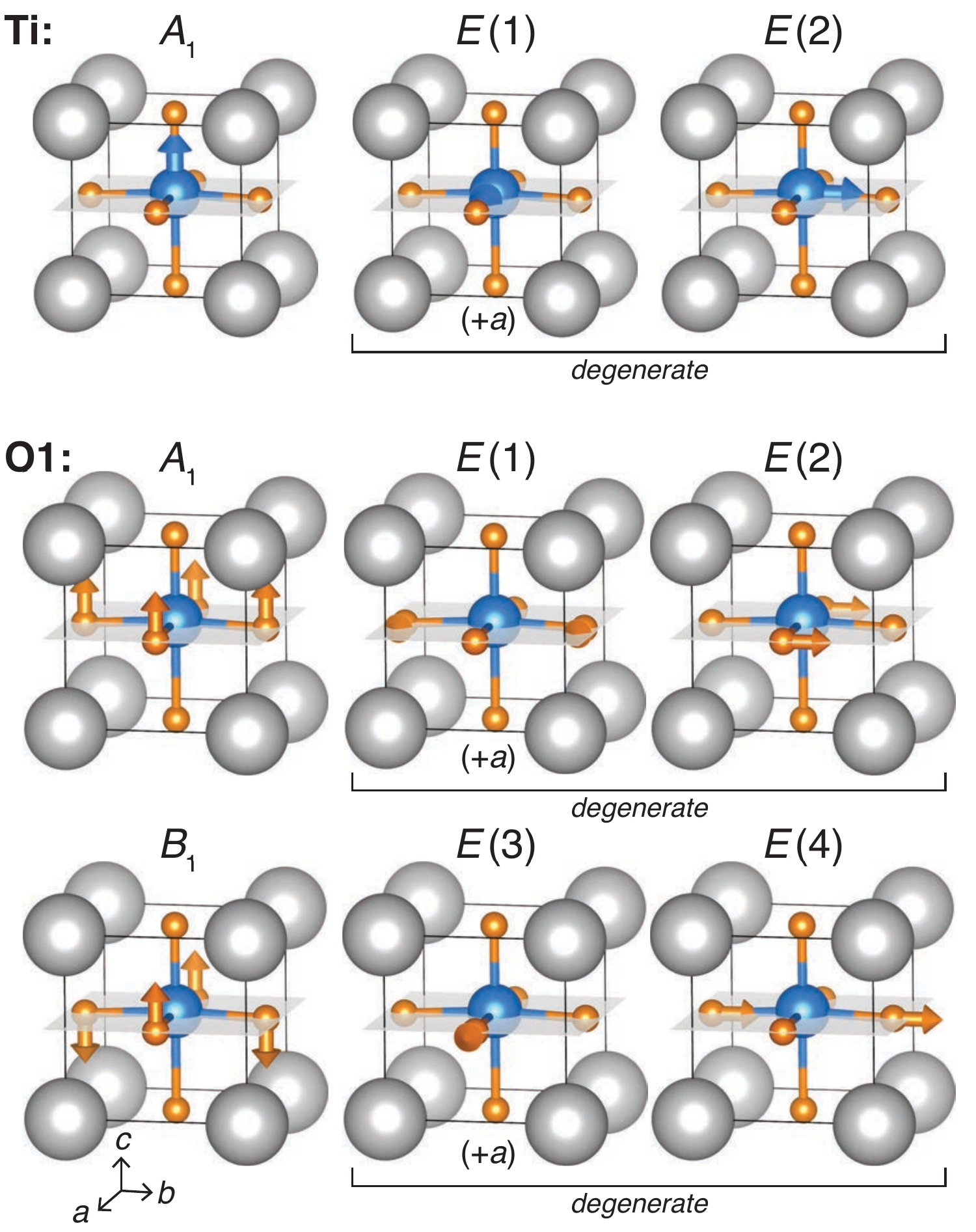}
\caption{Visualization of selected tight-binding modes of BaTiO$_3$ in the $P4mm$ space group.  For Ti ($1b$ Wyckoff position), all three basis vectors are shown and demonstrate the retained 3$N$ degrees of freedom as the pair that together transform as $E$ are allowed to have independent amplitudes.  For the O2 site ($2c$ Wyckoff position), the $A_1$ and $B_1$ modes each joins two oxygen positions, but $3N$ degrees of freedom are retained for the two atoms generated from that Wyckoff position, noted by the six independent modes.  While the $E$ pairs will transform together if the local symmetry is also $P4mm$, all amplitudes are allowed to vary independently in this analysis.   }
\label{fig:BaTiO3irreps}
\end{center}
\end{figure}

The ferroelectric ceramic BaTiO$_3$ at $T=298$~K provides an excellent example of a local distortion that averages out to a higher crystallographic symmetry in the unit cell.   The average crystallographic symmetry determined from Rietveld analysis is tetragonal, $P4mm$, which was used to define the tight-binding modes.  However, the local bonding environment is significantly distorted and better described by the symmetry of the low-temperature $R3m$ configuration \cite{kwei1995pair,ravel1998local,doi:10.1021/cm100440p}, as illustrated in Figure~\ref{fig:BaTiO3clouds}(a) and (b).  While the $R3m$ model provides a quantitative fit to the PDF within one unit cell, it does not provide quantitative information of the medium-range order, such as information of the correlations between unit cells or the coherence length scale, even though such information can (and should) exist within the PDF.

\begin{table}[h]
\caption{Basis vector components along each crystallographic direction for  BaTiO$_3$, described in the $P4mm$ space group setting, using the fractional atom coordinates, Ba: (0~0~0), Ti: (0.5~0.5~0.516), O1$a$: (0.5~0~0.487), O1$b$: (0~0.5~0.487), O2: (0.5~0.5~0.978).  }
\begin{center}
\begin{tabular}{ll|c|ccc}	
	& 	&		&	\multicolumn{3}{c}{Basis vector components}					\\
Irrep	&	&	Atom	&	$u||a$	&	$u||b$	&	$u||c$	\\
\hline										
\hline										
Ba	$A_{1}$	& &	Ba	&	0	&	0	&	1	\\
Ba	$E$    &(1)	&	Ba	&	1	&	0	&	0	\\
Ba	$E$&   (2)	&	Ba	&	0	&	1	&	0	\\
\hline										
Ti	$A_{1}$&	&	Ti	&	0	&	0	&	1	\\
Ti	$E$&(1)	&	Ti	&	1	&	0	&	0	\\
Ti	$E$&(2)	&	Ti	&	0	&	1	&	0	\\
\hline										
O1	$A_{1}$&	&	O1$a$	&	0	&	0	&	1	\\
	&	&	O1$b$	&	0	&	0	&	1	\\
O1	$B_1$&	&	O1$a$	&	0	&	0	&	-1	\\
	&	&	O1$b$	&	0	&	0	&	1	\\
O1	$E$&(1)	&	O1$a$	&	1	&	0	&	0	\\
	&(1)	&	O1$b$	&	0	&	0	&	0	\\
O1	$E$&(2)	&	O1$a$	&	0	&	0	&	0	\\
	&(2)	&	O1$b$	&	0	&	1	&	0	\\
O1	$E$&(3)	&	O1$a$	&	0	&	0	&	0	\\
	&(3)&	O1$b$	&	1	&	0	&	0	\\
O1	$E$&(4)	&	O1$a$	&	0	&	1	&	0	\\
	&(4)	&	O1$b$	&	0	&	0	&	0	\\
\hline										
O2	$A_{1}$&	&	O2	&	0	&	0	&	1	\\
O2	$E$&(1)	&	O2	&	1	&	0	&	0	\\
O2	$E$&(2)	&	O2	&	0	&	1	&	0	\\
\end{tabular}
\end{center}
\label{tab:BaTiO3irreps}
\end{table}%

\begin{figure}[t]
\begin{center}
\includegraphics[width=2.75in]{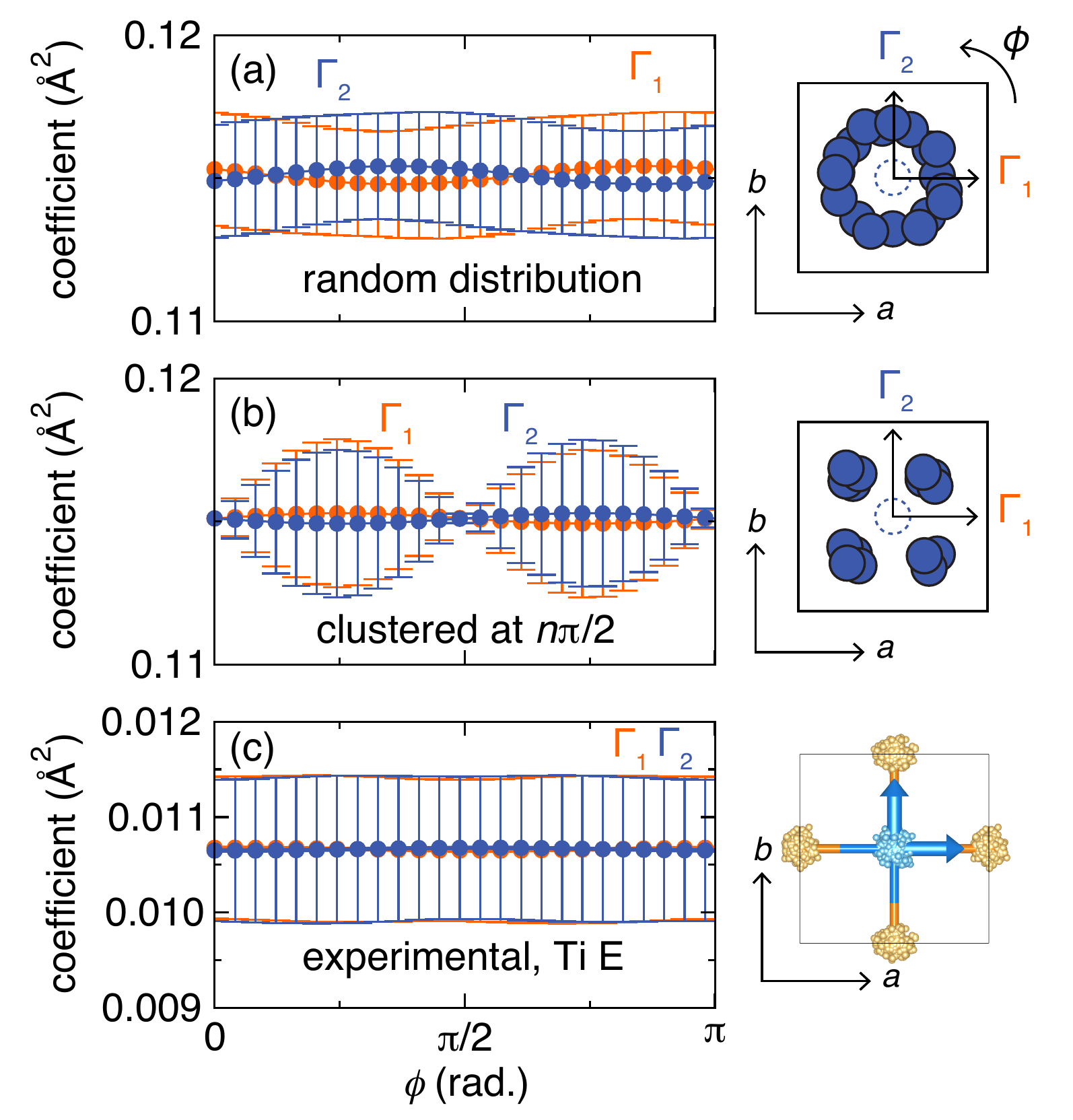}
\caption{ Euler angle analysis of a multidimensional tight-binding mode set that transforms as a multidimensional irreducible representation.  (a) A random distribution of atom positions from the crystallographic location  (dashed circle in cartoon) produces an equivalent variance of the mixing coefficients (denoted with vertical bars)  between the two basis vectors ($\psi_1$ and $\psi_2$) when the basis vectors are rotated about the Euler angle, $\phi$, parallel to the $C_4$ axis of the crystal structure.  (b) A clustering of  positions at regular intervals, such as $\pi/2$, will produce the same mixing coefficients as in (a) for each tight-binding mode when averaged over all $\vec{k}$ and over all ensembles.  However, a clustering of positions will yield a significant variance  in the coefficients, denoted by the vertical bars.    (c) The coefficients provided from the experimentally derived BaTiO$_3$ ensembles do not display significant differences when the basis vectors are rotated about the Euler angle.    }
\label{fig:euler}
\end{center}
\end{figure}

The experimental data used for this analysis were collected using the NPDF instrument (Lujan Neutron Scattering Center, Los Alamos National Laboratory) and were reanalyzed with adjusted relative absorption corrections (such that a scale factor was not needed to fit the intensity $G(r)$); the experimental details and original report of the experimental data can be found in Ref.~\cite{doi:10.1021/cm100440p}.  The Bragg profile and PDF were used to constrain RMC simulations using the RMCprofile code, \cite{Tucker:2007vt}  as illustrated in Figure~\ref{fig:outline}(a) and (b).   The simulation ensemble is a $12\times12\times12$ enlarged big-box ensemble of the tetragonal $P4mm$ unit cell (8640 atoms) that was determined from Rietveld analysis.   The ensembles were constrained by $G(r)$ (from $1<r<24$ \AA) in addition to the Bragg profile from the 90$^\circ$ detector bank  from NPDF ($ 1.7 < Q < 15.7$ \AA$^{-1}$;  $3.7> d > 0.4$ \AA).  In addition to hard-sphere cutoffs, a small penalty was applied to the simulations for breaking [TiO$_6$] coordination in order to accelerate the simulations.    200 different simulations were performed from the same starting configuration in order to build statistics in the analysis.  Each simulation ensemble can be back-folded into the unit cell; the atom positions fall within a cloud-like distribution centered around the average crystallographic site [Figure~\ref{fig:BaTiO3clouds}(c)].

Using the analysis method presented here, the atom positions were then decomposed into the tight-binding basis of the $P4mm$ space group with a $k$-mesh divided into 12 discrete steps along each $x$, $y$, and $z$ direction with $M_x=M_y=M_z= 12$.   The irreducible representations and corresponding basis vectors for the tight-binding (locally symmetry-adapted) modes were identified using the Bilbao Crystallographic Server (Symmetry Adapted Modes)\cite{doi:10.1080/0141159031000076110} and are listed in Table~\ref{tab:BaTiO3irreps}; some basis functions are graphically represented in Figure~\ref{fig:BaTiO3irreps}.

\begin{table}[h]
\scriptsize
\caption{
Amplitudes of the tight-binding modes (summed over all wave vectors) for the numerical average of 12 BaTiO$_3$ ensembles, a control simulation assuming $P4mm$ local symmetry, and a control simulation assuming $R3m$ local symmetry. The values in parenthesis are the variance of the coefficient. }
\label{tab:batio3coefs}
\begin{center}
\begin{tabular}{l|c|c|c}
			&	\multicolumn{3}{c}{Tight-binding mode coefficients (\AA$^2$)}							\\
Irrep			&	Data	&		$P4mm$ control	&	$R3m$ control	\\
\hline											
Ba	$A_{1}$	&	0.0034(1)	&	0.0059(4)	&	0.0026(3)	\\
Ba	$E$(1)	&	\textbf{0.0127(4)	}&	0.0066(5)	&	0.007(1)	\\
Ba	$E$(2)	&	\textbf{0.0130(3)	}&	0.0065(5)	&	0.007(1)	\\
Ti	$A_{1}$	&	0.0089(3)	&	0.0082(6)	&	0.0024(3)	\\
Ti	$E$(1)	&	\textbf{0.0143(6)	}&	0.0077(6)	&	0.009(1)	\\
Ti	$E$(2)	&	\textbf{0.0128(5)	}&	0.0076(6)	&	0.009(1)	\\
O1	$A_{1}$	&	0.0044(1)	&	0.0043(3)	&	0.0031(4)	\\
O1	$E$(1)	&	\textbf{0.0103(3)	}&	0.0068(5)	&	0.008(1)	\\
O1	$E$(2)	&	\textbf{0.0101(3)	}&	0.0068(5)	&	0.008(1)	\\
O2	$A_{1}$	&	0.0040(1)	&	0.0045(3)	&	0.0025(3)	\\
O2	$B_1$	&	0.0039(1)	&	0.0044(3)	&	0.0024(3)	\\
O2	$E$(1)	&	\textbf{0.0116(4)	}&	0.0076(6)	&	0.009(1)	\\
O2	$E$(2)	&	\textbf{0.0114(4)	}&	0.0076(6)	&	0.009(1)	\\
O2	$E$(3)	&	\textbf{0.0090(3)	}&	0.0066(5)	&	0.007(1)	\\
O2	$E$(4)	&	\textbf{0.0099(2)	}&	0.0066(5)	&	0.007(1)	\\
\end{tabular}
\end{center}
\label{default}
\end{table}%

For the analysis of a single ensemble of BaTiO$_3$, the tight-binding mode amplitudes that describe displacements along the ferroelectric polarization are not very large (Ti $A_{1}$, O1 $A_1$, O2 $A_1$, O2 $B_1$, Table~\ref{tab:batio3coefs}).  This makes sense, since the average positions of the titanium and O2 atoms  off-center along the elongated $c$ axis direction (Table~\ref{tab:BaTiO3irreps}) \cite{megaw1945crystal,Megaw}.  However, the displacements in the $ab$ plane are significantly enlarged.  This is graphically represented by the ``point-cloud'' distributions of the atom positions in Figure~\ref{fig:BaTiO3clouds}(c) that are overlain on top of the $R3m$ unit cell used to describe the PDF in Ref.~\cite{doi:10.1021/cm100440p}.  

One problem with RMC simulations is that if the data are insufficiently resolving such that some atoms are poorly constrained, then the simulation atoms can wander away from their ideal positions.  This would give the same graphical appearance as in Figure~\ref{fig:BaTiO3clouds}(c); however, the quantitative data presented in Table~\ref{tab:batio3coefs} shows that these displacements are significant on average within an ensemble and that their variance is tightly defined even across 200 simulations.

As a control, we performed RMC simulations constrained by \emph{simulated} pair distribution functions.  In one case ($P4mm$ control), we computed $G(r)$ from the $P4mm$ crystal structure obtained by Rietveld analysis (convoluted with the appropriate instrumental resolution parameters, $Q_{damp}$ and $Q_{broad}$); the Bragg profile was the experimental Bragg profile.  The simulated $G(r)$ and Bragg profile were used to constrain 200 RMC simulations for analysis.  For another control ($R3m$ control), we took the reported $R3m$ model determined from small-box modeling for the PDF (as reported in Ref.~\cite{doi:10.1021/cm100440p}) and simulated $G(r)$ from that structural model; the Bragg profile was the experimental Bragg profile).  These then constrained  90 independent RMC simulations for analysis.  The $P4mm$ control is a negative control that does not have additional displacements within the $ab$ plane (beyond thermal disorder modeled by a Debye-Waller factor); the $R3m$ control is a positive control for a known displacement in the $ab$ plane coincident with thermal disorder.  The tight-binding mode coefficients resulting from analysis of the $P4mm$ control simulations do not have a substantial anisotropy (Table~\ref{tab:batio3coefs}); while there is a statistically significant increase in the coefficients of displacements in the $ab$ plane, this may be biased from using the experimental Bragg peaks in conjunction with the simulated $G(r)$.  For the $R3m$ control, there is a significant and expected increase in displacements within the $ab$ plane.  This analysis informs us that the tight-binding mode amplitudes are capable of identifying aperiodic displacements when expected; however, the values of the coefficients alone do not inform us to if particular symmetry operations are broken.

\begin{figure}[t]
\begin{center}
\includegraphics[width=2.75in]{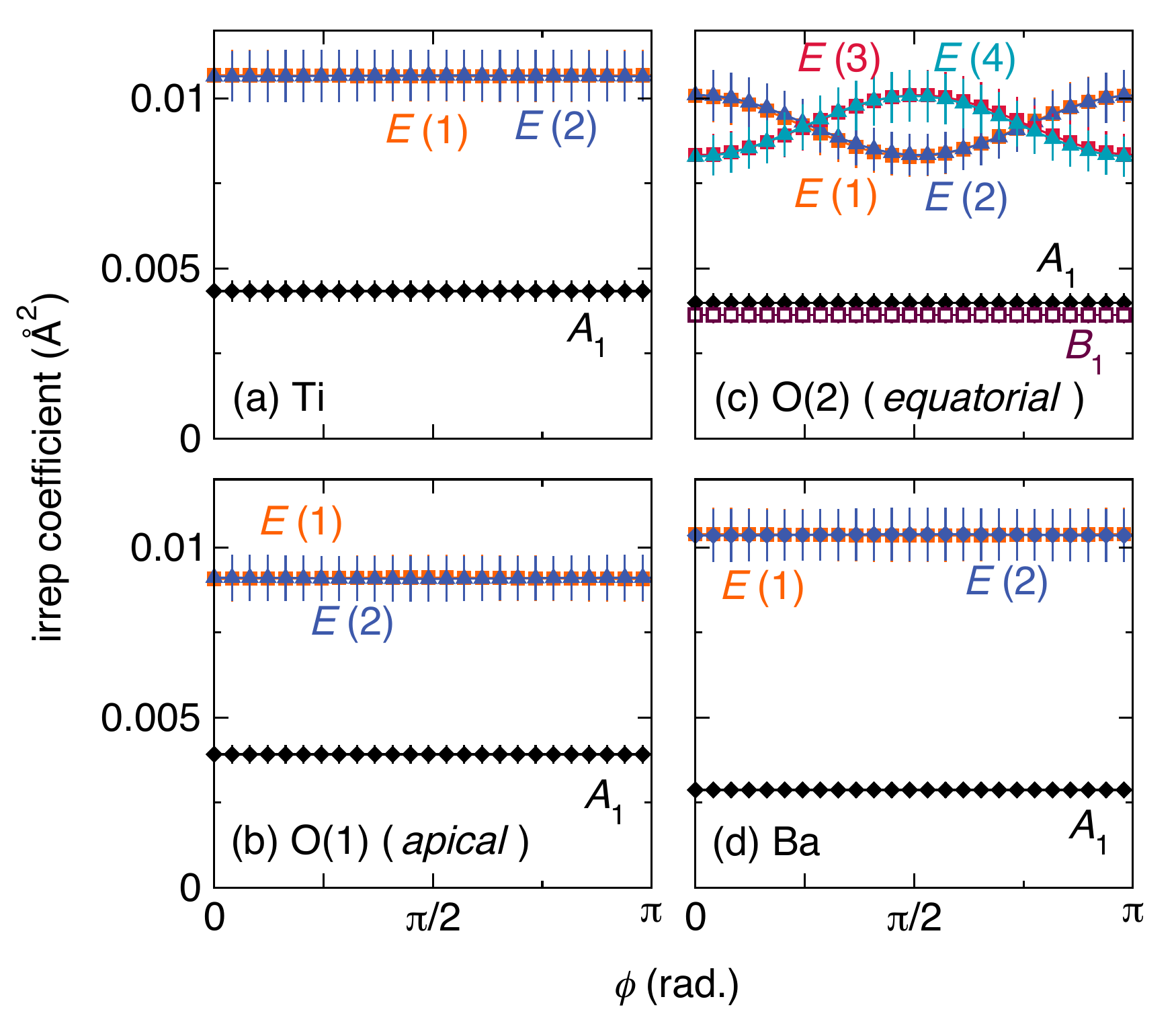}
\caption{Tight-binding mode coefficients as a function of Euler angle, $\phi$, parallel to the $C_4$ axis of $P4mm$ for (a) Ti, (b) O(1), (c), O(2), and (d) Ba crystallographic sites, illustrating a constant variance as a function of $\phi$.   }
\label{fig:eulerdata}
\end{center}
\end{figure}

With a local  trigonal distortion, the $R3m$-based model implies that there are specific vectors that the Ti displacements orient along -- these are the vectors that point directly at the faces of the [TiO$_6$] octahedra (i.e., the $\langle 111 \rangle$ directions, as referenced to the $P4mm$ or $Pm\bar{3}m$ unit cells of BaTiO$_3$).  However, by looking at the graphical representation in Figure~\ref{fig:BaTiO3clouds}(c), it is impossible to tell if particular directions are preferred.  
Because the tight-binding modes with a set are mutually orthogonal and therefore yield locally orthogonal displacements, it is trivial to rotate the reference frame of the basis vectors and recompute their coefficient as a function of the Euler angle  along the rotation axis of the multidimensional irreducible representation.  In the $P4mm$ description, this angle ($\phi$) rotates around the 4-fold axis of the unit cell.   

In our analysis, we decomposed the atomic displacements into amplitudes of specific tight-binding modes as a function of rotation about the Euler angle, $\phi$ (Figure~\ref{fig:euler}).  To illustrate this analysis, we employ two control simulations.  Shown in Figure~\ref{fig:euler}(a) is a simulation of displacements of Ti atoms around an approximately random distribution of $\phi$ angles.  In Figure~\ref{fig:euler}(b), we show a simulation with Ti atoms displaced at the same magnitude as in Figure~\ref{fig:euler}(a), but the angles are constrained to  be a random integer multiple of $\pi/2$ rad.  Therefore, the Ti atoms cluster into four groups (akin to the $\langle 111 \rangle$ displacements).  In both cases, the average coefficient of the tight-binding modes that together from a set and span a multidimensional irreducible representation will not change as a function of $\phi$, since the Ti atoms are displaced from the center by the same distance.  However, the variance between tight-binding mode amplitudes [$E(1)$ \emph{vs.} $E(2)$] will be distinct for each Euler angle (\emph{cf.}, Figure~\ref{fig:SchemeEuler}).    For the completely random distribution in Figure~\ref{fig:euler}(a), the coefficient multiplying $\Gamma_1$ of the $E$ irreducible representation will continuously vary between 0 and the maximum value, as the basis vector is orthogonal and collinear with the atomic displacement; the second basis vector ($\Gamma_2$) will also vary the same amount, but its amplitude will be $\pi/2$ out-of-phase with $\Gamma_1$.  Therefore, each basis vector will have the same variance with $\phi$, denoted by the error bars in Figure~\ref{fig:euler}.

\begin{figure}[t]
\begin{center}
\includegraphics[width=2.75in]{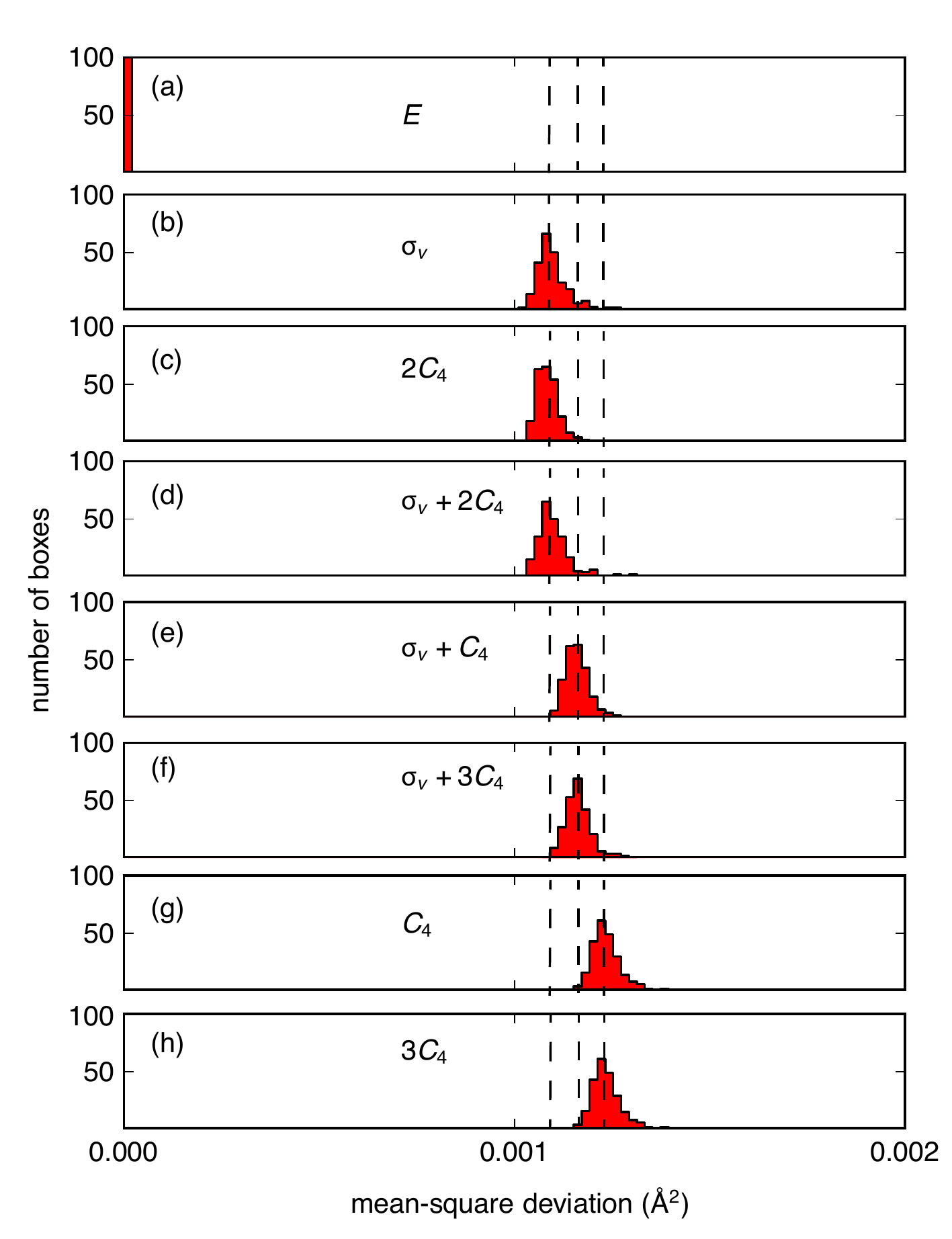}
\caption{Histograms of the mean-square displacements, summed over all $\vec{k}$ and mode sets ($j$) for 240 ensembles for each  symmetry operation of $P4mm$, (a) the identity, $E$, (b) a vertical mirror along $x$, $\sigma_v$, (c) a two-fold axis, $2C_4 = C_2$, (d) $\sigma_v+2C_4$, (e) a vertical mirror plane along $xy$, $\sigma_v+C_4 = \sigma_{xy}$,  (f)  the four-fold rotation axis, $C_4$, and (g) $3C_4$.    }
\label{fig:BaTiO3orients}
\end{center}
\end{figure}

As in Figure~\ref{fig:euler}(b), if the atom displacements cluster into groups, then the variation of basis vector coefficients will not be constant with $\phi$.  When $\phi$ is 0 rad., such that $\Gamma_1$ is oriented along the $a$ axis, then its mixing coefficient will be  $\sqrt{2}$ times the average value; the coefficient of $\Gamma_2$ will be identical.  Therefore, the difference in coefficients is zero.  However, when $\phi$ orients one of the basis vectors directly at the clustered displacements, one coefficient is maximal, and the other is zero; this produces a large variation in the basis vector amplitudes.  In the experimental simulations, there does not appear to be explicit clustering of the titanium atoms along particular displacement vectors [Figure~\ref{fig:euler}(c)].  Looking at the variation in coefficients for all atoms in the unit cell, depicted by the error bars in Figure~\ref{fig:eulerdata}, there does not appear to be any clustering of displacements as a function of Euler angle.  While the two-dimensional $E$ irreducible representations for O2 appear to exhibit a trend with $\phi$, the change in average value of the coefficient reflects the definition of the basis vectors; the variations of the coefficients, as indicated by the error bars, do not change with $\phi$.    This result is consistent for RMC simulations run for different times (as disorder tends to be artificially maximized for longer simulation runs).

\begin{figure}[t]
\begin{center}
\includegraphics[width=2.75in]{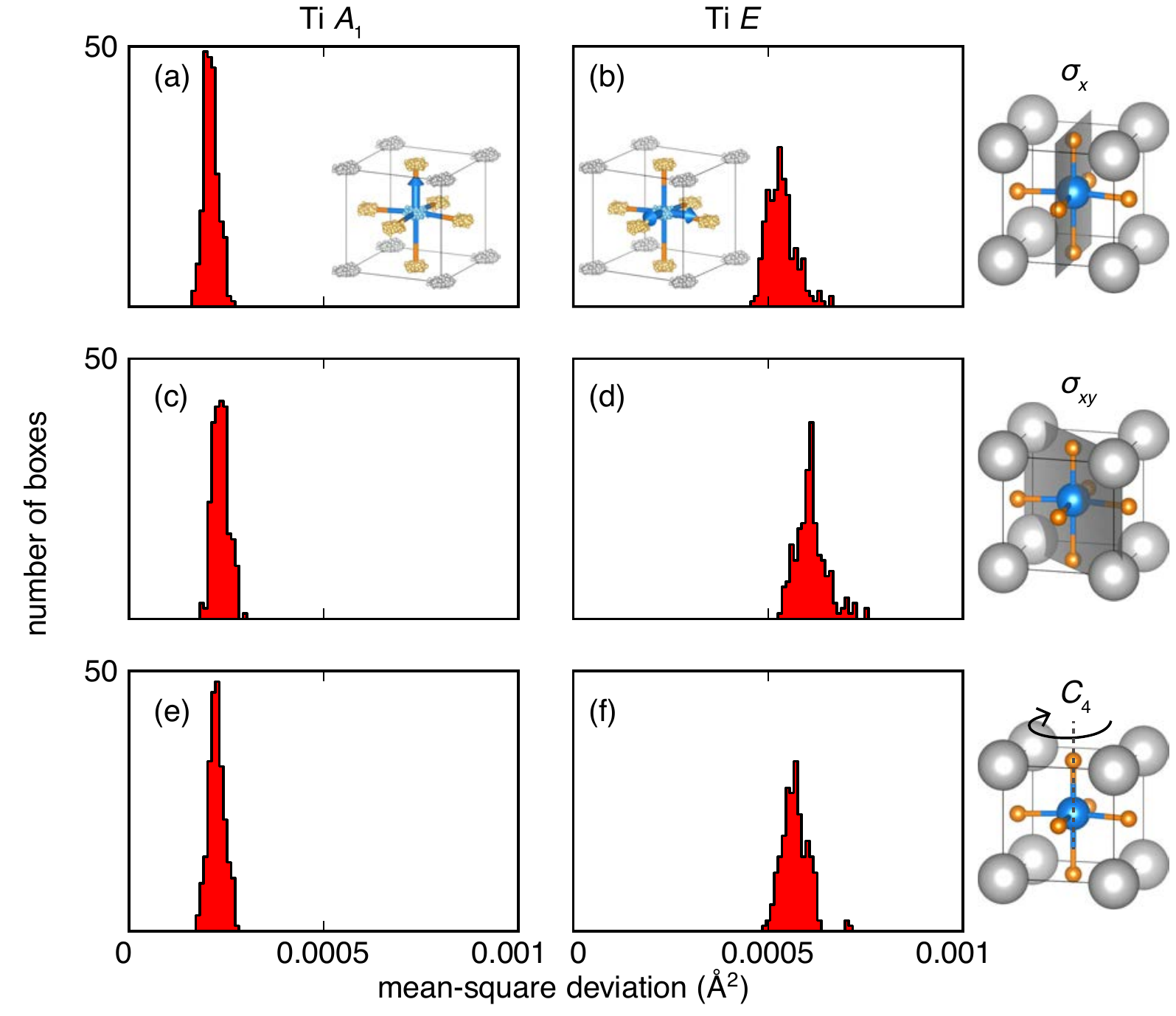}
\caption{Histograms of the mean-square displacements of the  (a, c, e) Ti $A_1$ and (b, d, f) Ti $E$ mode sets, as summed over all $\vec{k}$ for each unique symmetry orientation:  (a,b) $\sigma_x$ mirror planes, (c,d) $\sigma_{xy}$ mirror planes, and (e,f) $C_4$ rotation axis.   }
\label{fig:Tiorients}
\end{center}
\end{figure}

While a variation of coefficients with Euler angle can indicate clustering of displacements described by a multidimensional irreducible representation, it does not provide any indication if the degeneracy-inducing symmetry operation is broken.  To find broken degeneracies, we turn to continuous symmetry measures as defined in the methods.  For BaTiO$_3$, we compute the mean-square deviation (MSD) for each generated symmetry operation of the crystallographic space group ($P4mm$).  With four symmetry operations ($E$, $\sigma_v$, $C_2$, and $C_4$), there are a total of 8 symmetry-related atoms that are generated from a general position; therefore, we test all unique combinations of these operations (each combination that generates one of the general positions).

The MSDs illustrate that the atomic displacements in the ensembles show the highest deviation away from four-fold rotation symmetry element.  Histograms of all MSDs computed for BaTiO$_3$ (summed over all $\vec{k}$ and irreducible representations) are illustrated in Figure~\ref{fig:BaTiO3orients} for each symmetry operation.  The histograms for related symmetry elements cluster together; those combinations that equate to a 4-fold rotation have the most significant MSD [Figure~\ref{fig:BaTiO3orients}(g,h)], followed by mirror planes parallel to the $\{110\}$ planes, then mirror planes parallel to the $\{100\}$ planes.

To probe which irreducible representations are most symmetry conserving, histograms of MSDs summed over all $\vec{k}$ for each irreducible representation are shown in Figure~\ref{fig:Tiorients}; the histograms bin together MSDs computed for the equivalent symmetry operations shown on the right.  The histograms for the Ti $A_1$ irreducible representation [Figure~\ref{fig:Tiorients}(a,c,e)] show tightly-grouped and low-valued MSDs indicating that the vertical Ti displacements tend to preserve the $P4mm$ symmetry operations.  However, the displacements that project onto the Ti $E$ irreducible representation tends to break symmetry operations, as previously inferred.  The four-fold rotation axis appears to be the symmetry operation most broken, as naively expected from the small-box trigonal model illustrated in Figure~\ref{fig:BaTiO3clouds}(b), which does have a vertical mirror plane parallel to the $(110)$ plane.

The analyses presented here for BaTiO$_3$ provide results that are sufficiently simple as to be easily compared with small-box models of BaTiO$_3$.  The use of RMC simulations allows one to extract a single, statistically relevant model of the atom positions that describes both the data regarding  local atom separations (the PDF) and  average crystallographic symmetry (Bragg profile).  For BaTiO$_3$, the coefficients of the tight-binding modes and their spatial dependence reveal the presence of a significant distortion from the $P4mm$ crystallographic symmetry.  The resulting ensemble reveals that the atom positions are mostly displaced in the $ab$ plane, which closely resembles the low-temperature $R3m$ crystal structure.  This example illustrates how such an analysis may be performed on materials  with more  complexity, both in terms of the crystal structure and the crystalline disorder, as described in the next section.

\subsection{Correlated Oxygen and Bismuth Displacements in Bi$_2$Ti$_2$O$_7$}

The analysis methods presented here are generally applicable to materials with more complex structures.  The ``charge-ice'' pyrochlore oxide, Bi$_2$Ti$_2$O$_7$, has a  large unit cell that contains 88 atoms; direct inspection of ensembles becomes prohibitive with this many degrees-of-freedom in a single crystallographic unit cell \cite{hector2004synthesis}.  In Bi$_2$Ti$_2$O$_7$, there is extensive disorder of the Bi sublattice, as attributed to stereochemical activity of the lone pair -- derived from the [Xe]$5d^{10}$6$s^2$ electron configuration of Bi(III) -- on a geometrically frustrated lattice.  The geometry of the diamond lattice prevents long-range ordering of the dipoles, in a manner related to Pauling's ice rules \cite{seshadri2006lone}.  Previously, RMC simulations of total neutron scattering have been used to gain an atomistic representation of the static Bi displacements, which form a toroidal distribution of Bi atoms that encircle the ideal crystallographic site.  Furthermore, the O$^\prime$ atoms  (8$a$ Wyckoff site) are connected to the non-spherically distributed Bi atoms and therefore become pulled off of their ideal crystallographic site into  tetrahedral volumes.    The original report, experimental data, and experimental details can be found in Ref.~\cite{Shoemaker_PRB_2010}.  The crystallographic Bi$_2$Ti$_2$O$_7$ unit cell is described by the $Fd\bar{3}m$ space group, which defines the irreducible representations and tight-binding modes used here.

\begin{figure}[t]
\begin{center}
\includegraphics[width=2.75in]{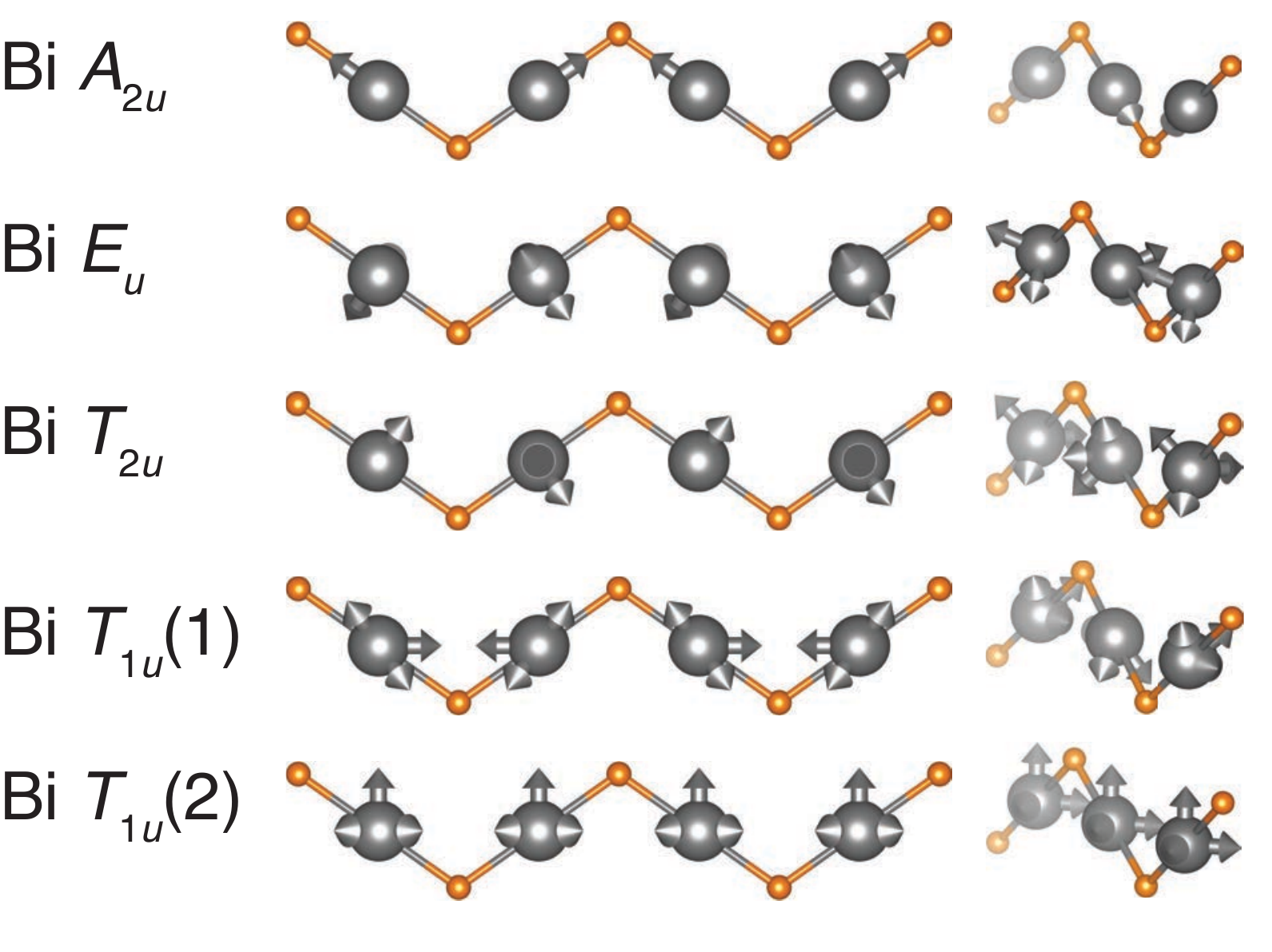}
\caption{(a) Visualization of the tight-binding modes for the different Bi irreducible representations.  The $E_u$, $T_{2u}$, and $T_{1u}(1)$ mode sets have the largest amplitude, which describe the displacements that generate toroidal distribution of Bi positions and agree with the predicted displacement modes of $E_u$ and $T_{1u}$ symmetry from \emph{ab initio} density functional theory calculations. \cite{Shoemaker_PRB_2010}.  }
\label{fig:bimodes}
\end{center}
\end{figure}

By rewriting the atomic displacements in terms of the tight-binding modes, a straightforward examination of their amplitudes reveals several characteristics that lead to many of the same conclusions as presented in Ref.~\cite{Shoemaker_PRB_2010}; these coefficients are tabulated in Table~\ref{tab:bi2ti2o7coefs} (the coefficients are averaged across modes related by face-centering, over all $\vec{k}$, and across 320 distinct ensembles).  
Of the Bi modes (depicted in Figure~\ref{fig:bimodes}), those spanning the  $E_u$ and $T_{2u}$ irreducible representations generate displacements that reproduce the toroidal distribution of Bi positions observed in Ref.~\cite{Shoemaker_PRB_2010} and have the most significant amplitude.  The modes spanning the $A_{2u}$ representation is orthogonal to the $C_\infty$ rotational axis of the torus and has a small amplitude.  The modes spanning the $T_{1u}(1)$ and $T_{1u}(2)$ representations have intermediate orientations and amplitudes.  The decomposition of atomic displacements into tight-binding modes reproduces the physically meaningful and intuitive results presented in Ref.~\cite{Shoemaker_PRB_2010}; here, the averaging across many wavevectors and simulations identifies the robustness of these displacements.  

Furthermore, identification of these high-amplitude modes allows one to create a ``small-box'' model for a symmetry constrained refinement.  In Refs.~\cite{Shoemaker_PRB_2010} and \cite{fennie2007lattice},  imaginary phonon modes were discovered at the Brillioun Zone center; the symmetry of the polarization eigenvectors belong to the $T_{1u}$ and $E_u$ irreducible representations.  The tight-binding modes spanning these irreducible representations have high-amplitude coefficients in the analysis performed here.  Distortion of the $Fd\bar{3}m$ lattice along these polarization modes yields a small unit cell of $Cm$ symmetry that provides an excellent description of $G(r)$ for $r<3.5$ \AA\ \cite{Shoemaker_PRB_2010}.  The agreement of the high-amplitude tight-binding modes with the theory-predicted distortion modes and small-box refinement illustrates another utility of this approach for unknown systems.

\begin{table*}[t]
\scriptsize
\caption{Tight-binding mode amplitudes obtained from averaging 320 RMC simulations of Bi$_2$Ti$_2$O$_7$.  Atom positions are defined as Bi: (0,0,0), Ti: (0.5, 0.5, 0.5), O1: (0.43114, 0.125, 0.125), O2: (0.125, 0.125, 0.125);  amplitudes are also averaged across atoms that are related through face-centering. \label{tab:bi2ti2o7coefs}}
\begin{center}
\begin{tabular}{ll|ll|ll|ll|ll|ll}
\hline
Irrep & Coef.~(\AA$^2$) & Irrep & Coef.~(\AA$^2$) & Irrep & Coef.~(\AA$^2$) & Irrep & Coef.~(\AA$^2$) & Irrep & Coef.~(\AA$^2$) & Irrep & Coef.~(\AA$^2$) \\
\hline
Bi 	$A_{2u}$	&	0.010(1)	&	Ti 	$A_{2u}$	&	0.003(4)	&		O1	$A_{1g}$	&	0.021(2)	&	O1	$T_{2g}$1(1)	&	0.007(1)	&	O1	$T_{1u}$2(1)	&	0.006(1)	&	O2	$T_{2g}$(1)	&	0.029	(3)	\\
Bi 	$E_u$(1)	&	0.067(8)	&	Ti 	$E_u$(1)	&	0.008(1)	&		O1	$A_{2u}$	&	0.006(1)	&	O1	$T_{2g}$1(2)	&	0.007(1)	&	O1	$T_{1u}$2(2)	&	0.006(1)	&	O2	$T_{2g}$(2)	&	0.029	(3)	\\
Bi 	$E_u$(2)	&	0.067(7)	&	Ti 	$E_u$(2)	&	0.008(1)	&		O1	$E_u$1	&	0.006(1)	&	O1	$T_{2g}$1(3)	&	0.007(1)	&	O1	$T_{1u}$2(3)	&	0.006(1)	&	O2	$T_{2g}$(3)	&	0.029	(3)	\\
Bi 	$T_{2u}$(1)	&	0.068(8)	&	Ti 	$T_{2u}$(1)	&	0.008(1)	&		O1	$E_u$2	&	0.006(1)	&	O1	$T_{2g}$2(1)	&	0.006(1)	&	O1	$T_{1u}$3(1)	&	0.007(1)	&	O2	$T_{1u}$(1)	&	0.030	(3)	\\
Bi 	$T_{2u}$(2)	&	0.067(8)	&	Ti 	$T_{2u}$(2)	&	0.008(1)	&		O1	$E_{g}$	&	0.006(1)	&	O1	$T_{2g}$2(2)	&	0.006(1)	&	O1	$T_{1u}$3(2)	&	0.007(1)	&	O2	$T_{1u}$(2)	&	0.030	(3)	\\
Bi 	$T_{2u}$(3)	&	0.067(8)	&	Ti 	$T_{2u}$(3)	&	0.008(1)	&		O1	$E_{g}$	&	0.006(1)	&	O1	$T_{2g}$2(3)	&	0.006(1)	&	O1	$T_{1u}$3(3)	&	0.007(1)	&	O2	$T_{1u}$(3)	&	0.030	(4)	\\
Bi 	$T_{1u}$1(1)	&	0.030(4)	&	Ti 	$T_{1u}$1(1)	&	0.005(1)	&		O1	$T_{2u}$1(1)	&	0.007(1)	&	O1	$T_{2g}$3(1)	&	0.007(1)	&	O1	$T_{1g}$1(1)	&	0.007(1)	&			&						\\
Bi 	$T_{1u}$1(2)	&	0.030(4)	&	Ti 	$T_{1u}$1(2)	&	0.005(1)	&		O1	$T_{2u}$1(2)	&	0.007(1)	&	O1	$T_{2g}$3(2)	&	0.007(1)	&	O1	$T_{1g}$1(2)	&	0.007(1)	&			&						\\
Bi 	$T_{1u}$1(3)	&	0.029(3)	&	Ti 	$T_{1u}$1(3)	&	0.005(1)	&		O1	$T_{2u}$1(3)	&	0.008(1)	&	O1	$T_{2g}$3(3)	&	0.007(1)	&	O1	$T_{1g}$1(3)	&	0.007(1)	&			&						\\
Bi 	$T_{1u}$2(1)	&	0.049(6)	&	Ti 	$T_{1u}$2(1)	&	0.006(1)	&		O1	$T_{2u}$2(1)	&	0.007(1)	&	O1	$T_{1u}$1(1)	&	0.008(1)	&	O1	$T_{1g}$2(1)	&	0.007(1)	&			&						\\
Bi 	$T_{1u}$2(2)	&	0.049(6)	&	Ti 	$T_{1u}$2(2)	&	0.006(1)	&		O1	$T_{2u}$2(2)	&	0.007(1)	&	O1	$T_{1u}$1(2)	&	0.008(1)	&	O1	$T_{1g}$2(2)	&	0.007(1)	&			&						\\
Bi 	$T_{1u}$2(3)	&	0.048(6)	&	Ti 	$T_{1u}$2(3)	&	0.006(1)	&		O1	$T_{2u}$2(3)	&	0.007(1)	&	O1	$T_{1u}$1(3)	&	0.008(1)	&	O1	$T_{1g}$2(3)	&	0.007(1)	&			&						\\
\hline
\end{tabular}
\end{center}
\end{table*}

\begin{figure}[t]
\begin{center}
\includegraphics[width=2.75in]{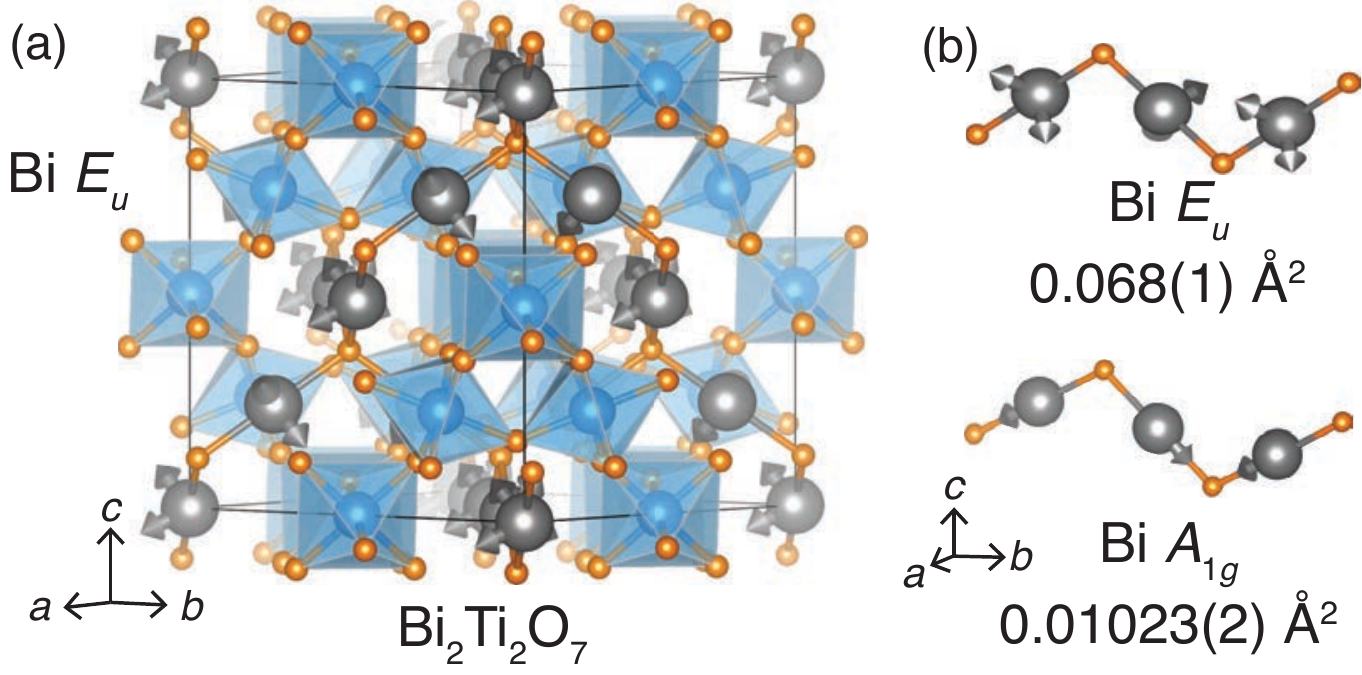}
\caption{(a) Visualization of tight-binding modes corresponding to the Bi $T_{2u}$ modes of Bi$_2$Ti$_2$O$_7$.  (b) The Bi $T_{2u}$ mode set shows the Bi atom displacing from its nominal 180$^\circ$ O$^\prime$--Bi--O$^\prime$ angle, orthogonal to the linear axis; this is reflected in the small magnitude of the Bi $A_{2u}$-derived modes.  }
\label{fig:BiT2u}
\end{center}
\end{figure}

\begin{figure}[t]
\begin{center}
\includegraphics[width=2.75in]{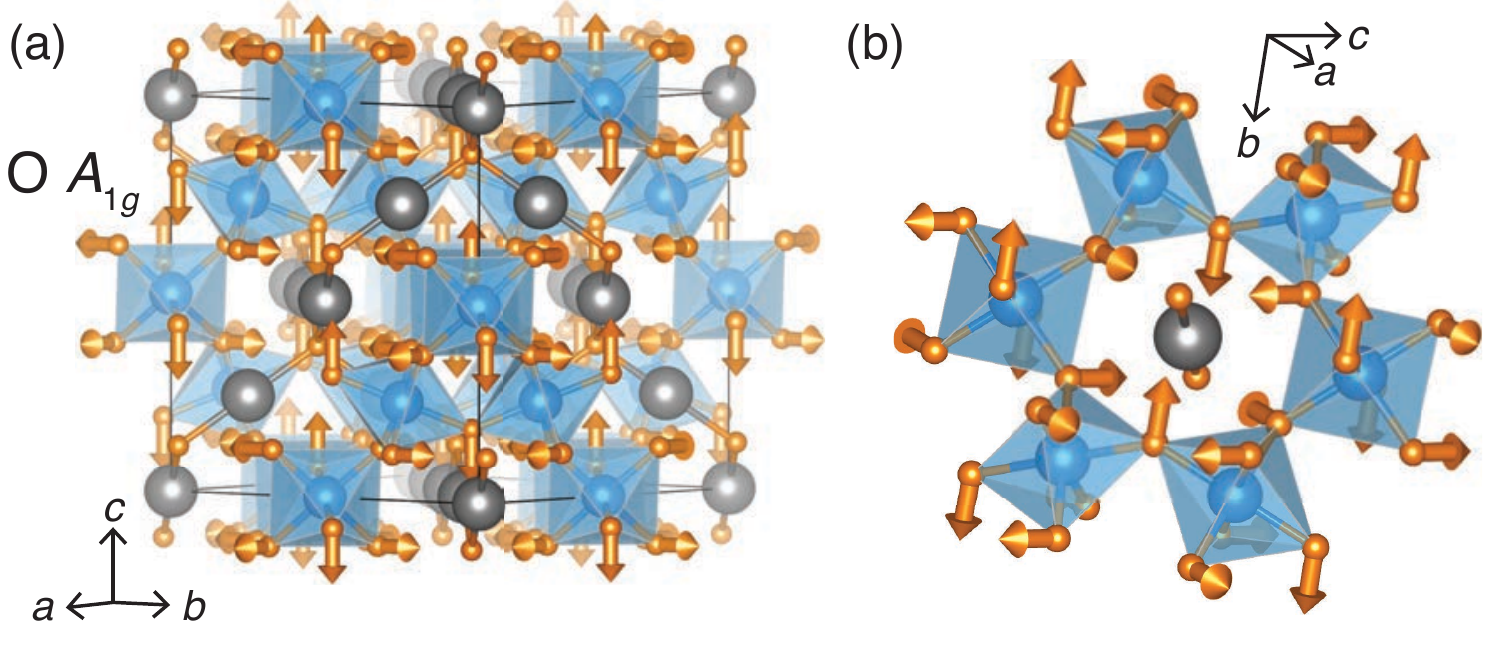}
\caption{(a) The O A$_{1g}$ representation carries the largest magnitude of all degrees of freedom for the  $48f$ oxygen atoms.  (b)  Vectors depicting the O $A_{1g}$ mode surrounding an linear [O$^\prime$--Bi--O$^\prime$] group illustrates how the oxygens move out of the plane in which the modes of the Bi $T_{2u}$ representation act (as illustrated in Figure~\ref{fig:BiT2u}).   While all vectors are illustrated, not all of the atoms shown are spanned by a single irreducible representation (i.e., those related by face-centering are included).   }
\label{fig:BiOmode}
\end{center}
\end{figure}

Additionally, the representational analysis performed here suggests that there are correlated Bi as well as O, Bi and O displacements that are not immediately observed from direct inspection of the atomic displacements.  While a  correlation between the Bi and O$^\prime$  displacements was previously made \cite{Shoemaker_PRB_2010}, the tight-binding mode amplitudes show that there are large displacements of Bi and O.  The highest modes corresponding to the 48$f$ oxygen atom correspond to the  O $A_{1g}$ irreducible representation, which can be described as a subtle elongation and twisting of the [TiO$_6$] octahedra [Figure~\ref{fig:BiOmode}(a)].  This large displacement is also mirrored in the anisotropic atomic displacement parameter of the 48$f$ oxygen position obtained from Rietveld analysis.   A possible origin of the high amplitude of this distortion is illustrated in Figure~\ref{fig:BiOmode}(b): the 48$f$ O atom positions form a hexagon encircling the linear O$^\prime$-Bi-O$^\prime$ linkages in an orthogonal orientation.  With significant Bi displacements, as indicated by the large amplitude of the Bi $T_{2u}$ spanning modes, the oxygen atoms displace from their ideal positions around the hexagon in order to accommodate the shifted Bi atoms.  

\begin{figure}[t]
\begin{center}
\includegraphics[width=2.75in]{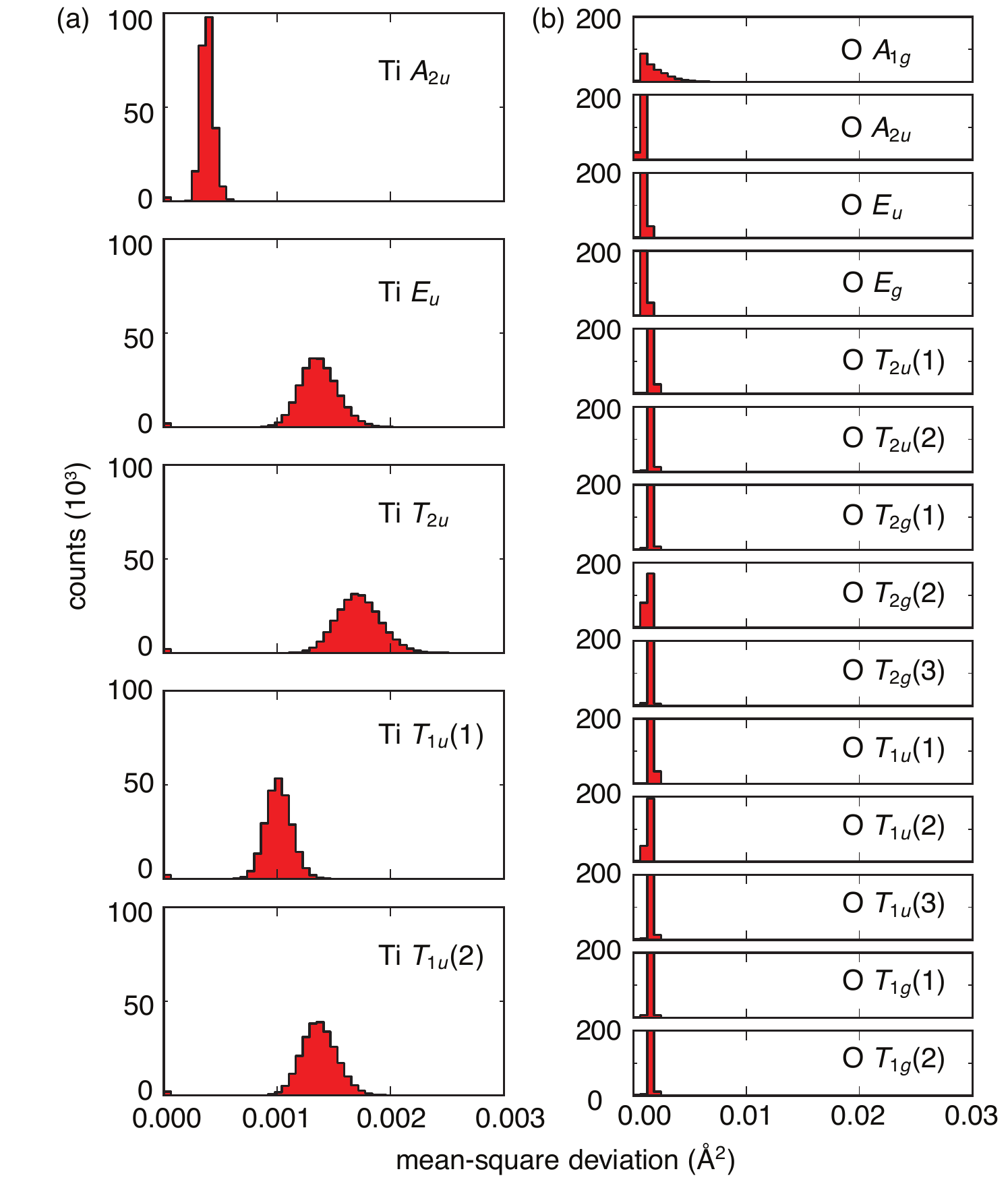}
\caption{Histograms of the mean-square displacements of the  (a)  Ti  and (b) O mode sets, as summed over all $\vec{k}$, all symmetry operations, and 320 ensembles illustrate that the Ti-O sublattice is not disrupted.   Only the O $A_{1g}$ mode shows any distribution of the MSD.    }
\label{fig:TiandOMSDs}
\end{center}
\end{figure}

By comparing different simulation runs, it is possible to gauge the uncertainty in how distorted or ideal the connectivity is in different parts of the lattice.  For example, the mode amplitudes corresponding to the Ti-O sublattice are shown in Figure~\ref{fig:TiandOMSDs}.  The histograms show that the displacement amplitudes have a narrow distribution across all length scales and between many simulation runs.  Furthermore, the displacements appear to be reasonably isotropic, as consistent with thermal disorder of a cubic lattice.    

\begin{figure}[t]
\begin{center}
\includegraphics[width=2.75in]{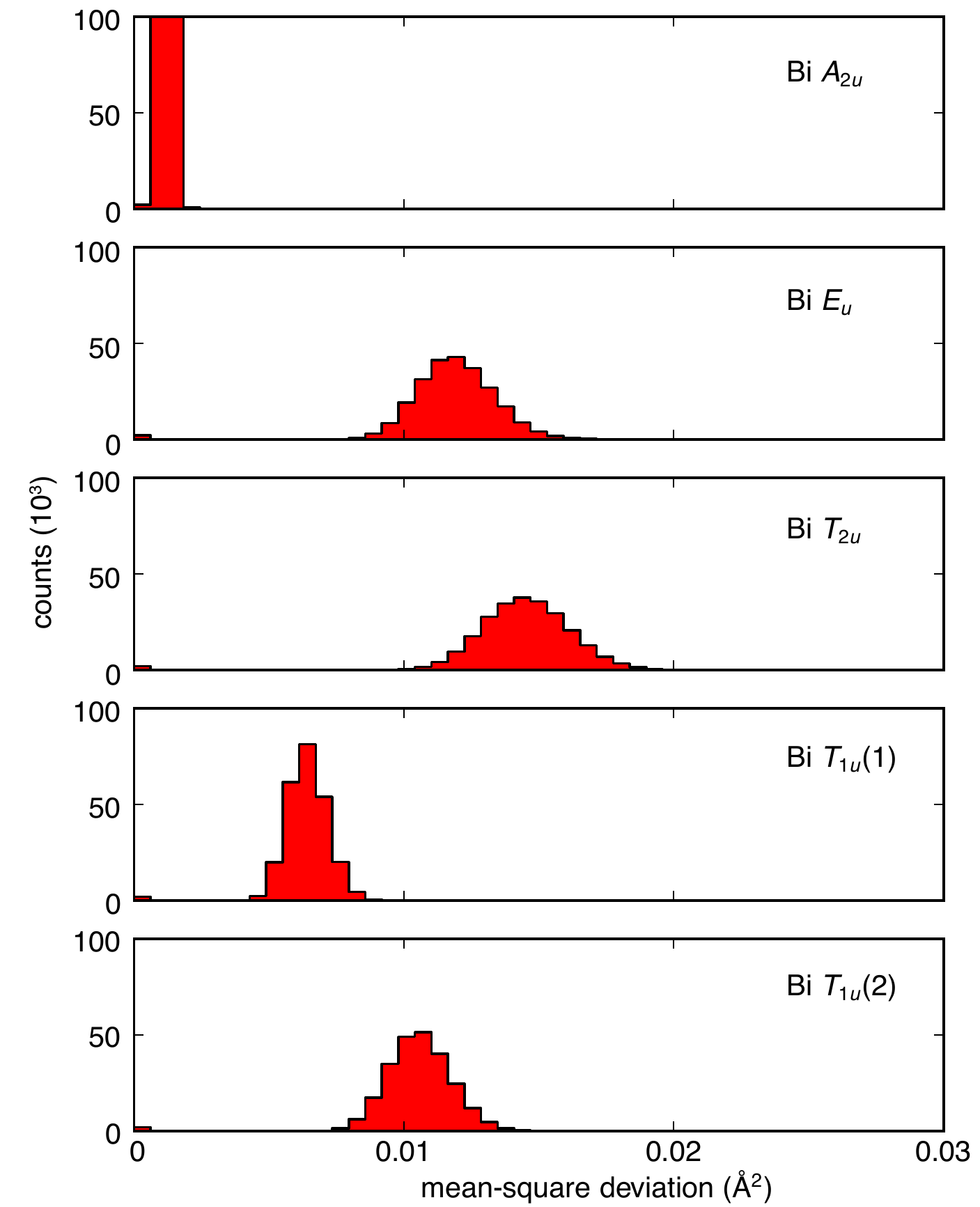}
\caption{Histograms of the mean-square displacements of the Bi mode sets, as summed over all $\vec{k}$, all symmetry operations, and 320 ensembles,  illustrate that the $E_u$, $T_{2u}$ and $T_{1u}(2)$ representations break the symmetry operations of $Fd\bar{3}m$; those modes correspond to toroidal displacements shown in Figure~\ref{fig:bimodes}.   }
\label{fig:BiMSDs}
\end{center}
\end{figure}

In contrast, the distribution of Bi atom displacements are varied (Figure~\ref{fig:BiMSDs}).  The Bi $A_{2u}$ mode distribution is comparable to the Ti-O sublattice.  However, many of the Bi displacement modes corresponding to multidimensional irreducible representations have high amplitudes and broad distributions, indicative of substantial static disorder in directions orthogonal to the linear O$^\prime$-Bi-O$^\prime$ bond axis.  This strongly suggests that those displacement modes locally break the $Fd\bar{3}m$ symmetry of the crystal structure.  

\begin{figure*}[t]
\begin{center}
\includegraphics[width=6.5in]{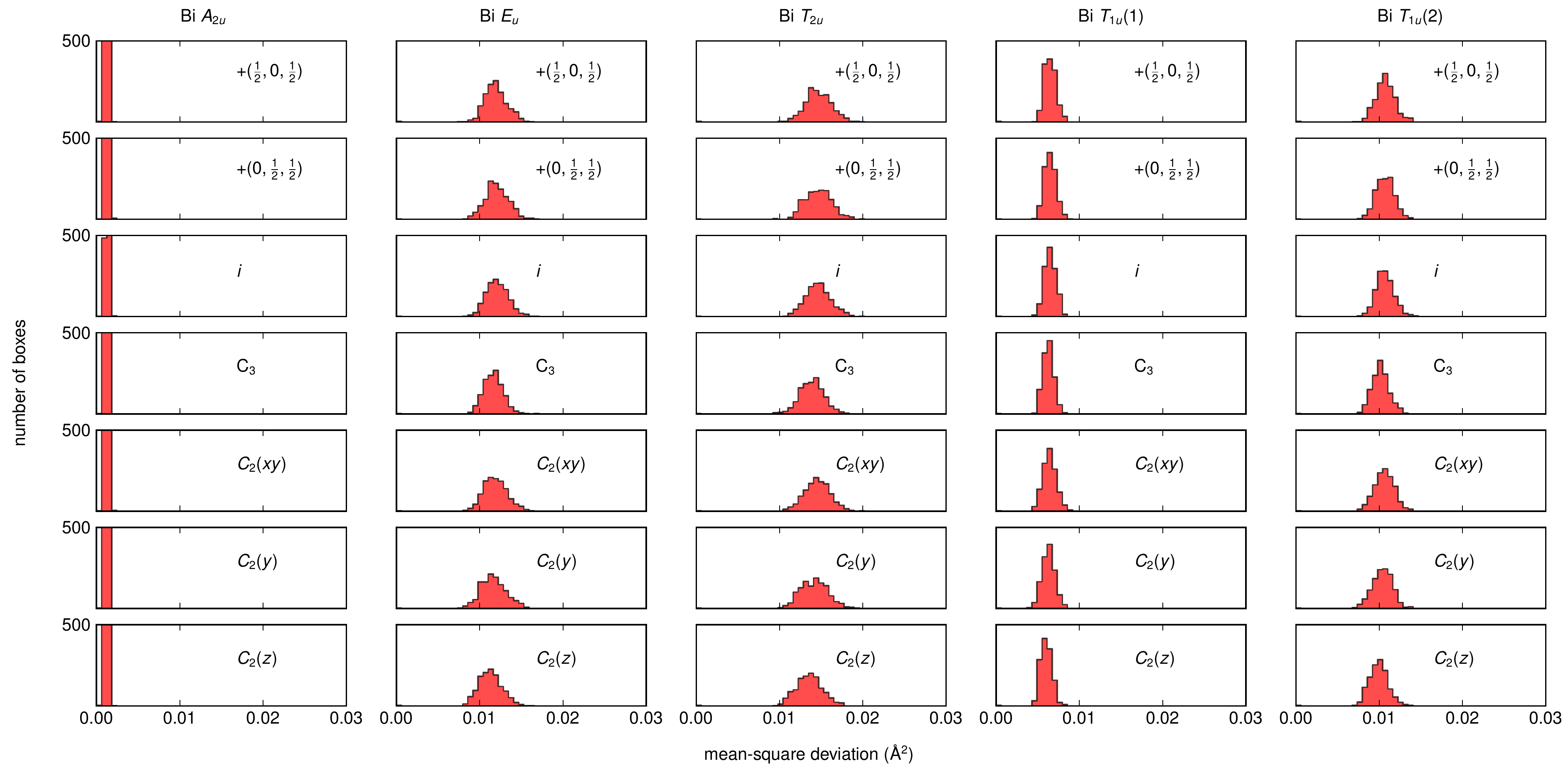}
\caption{Histograms of the mean-square displacements of the Bi mode sets for each symmetry operation of $Fd\bar{3}m$, as summed over all $\vec{k}$ and 320 ensembles.  The amplitude and breadth of the MSD's show that the  face centering [$+(\frac{1}{2},0,\frac{1}{2})$, $+(0,\frac{1}{2},\frac{1}{2})$] and inversion ($i$) symmetry operations are not well followed by the $E_u$, $T_{2u}$ and $T_{1u}(2)$ representations.  }
\label{fig:BiBySymmMSDs}
\end{center}
\end{figure*}

In this analysis, the multidimensional irreducible representations  are broken into their individual components (as to retain the total number of degrees of freedom); however, the values of each tight-binding mode are identical in the case of Bi$_2$Ti$_2$O$_7$.  To then identify if and by how much atomic displacements break the symmetry elements linking together tight-binding modes spanning a multidimensional irreducible representation,  the continuous symmetry measure of each irreducible representation  can be calculated.  
Figure~\ref{fig:BiBySymmMSDs} contains histograms of the MSDs for each irreducible representation after operation on the simulation box by a specific symmetry operation (Eqn.~\ref{eqn:csmirrep}).  The modes spanning the $A_{2u}$ representation do not show any dependence on the symmetry operation; however, the modes corresponding to $E_u$ and $T_{2u}$ representations do show a dependence on the operations.  Specifically, the face-centering [$+(\frac{1}{2},0,\frac{1}{2})$, $+(0,\frac{1}{2},\frac{1}{2})$] and inversion ($i$) symmetry operations show the highest MSDs, as well as the broadest distributions suggesting that those symmetries are deviated by the largest magnitude and in the most ways.  In future work, it will be informative to analyze the compatibility relationships as the degeneracy of different modes changes as $\vec{k} \neq (0,0,0)$.

The crystal structure of Bi$_2$Ti$_2$O$_7$ presents a very complex problem, as the unit cell contains 88 atoms, resulting in 264 degrees of freedom, or 264 distinct tight-binding modes, to describe all atom displacements.  When trying to analyze a large ensemble simulation of this structure, analysis in Cartesian coordinates becomes unwieldy.  Decomposition of the structure into the crystallographically relevant local basis allows one to determine the highest amplitude disorder in the lattice, the distribution of amplitudes, the direction of the atomic displacements causing the disorder,  how the disorder breaks specific symmetry elements of the crystallographic space group, and by how much.

\section{Conclusions}
The representational analysis of large atomistic ensembles generated from simulations (such as from reverse Monte Carlo simulations of total scattering data) using a tight-binding basis derived from locally symmetry-adapted modes  is a robust method that allows one to quantify disorder in the lattice.  In many RMC simulations, the goal is often to characterize subtle deviations from the lattice -- these types of displacements are subtle perturbations from a lattice that posses a modicum of moderately isotropic thermal disorder.  Therefore, isolation and quantification of the disorder (i.e., of infrequent events) requires statistical analysis.  By representing the disorder with respect to a local basis of the background signal (\emph{i.e.}, symmetry-adapted modes of the crystallographic space group), displacements appear as a positive signal, are amplified and can be statistically analyzed.   
 Additionally, the approach presented here permits a framework for analyzing other types of degrees of freedom, such as  occupational/compositional disorder (e.g., solid-solutions) or magnetism.  Such a rigorous group theoretical treatment is currently implemented in ISODISPLACE \cite{isodisplace}.  

\section{Acknowledgements}
The source code used in this analysis is freely available at: \href{https://occamy.chemistry.jhu.edu/references/pubsoft/index.php}{https://occamy.chemistry.jhu.edu/references/\linebreak{}pubsoft/index.php}.  The authors thank D. P. Shoemaker, K. Page, and R. Seshadri for sharing their neutron scattering data for this analysis and for helpful discussions.  
This research is principally supported by the US DoE, Office of Basic Energy Sciences (BES), Division of Materials Sciences and Engineering under Award DE-FG02- 08ER46544. TMM acknowledges support from the David and Lucile Packard Foundation.   This work benefited from the use of NPDF at the Lujan Center at Los Alamos Neutron Science Center, funded by DoE BES. Los Alamos National Laboratory is operated by Los Alamos National Security LLC under DoE Contract No. DE-AC52-06NA25396. The upgrade of NPDF was funded by the National Science Foundation through Grant No. DMR 00-76488. This research utilized the CSU ISTeC Cray HPC System supported by NSF Grant CNS-0923386.


\bibliographystyle{unsrt}

\end{document}